\documentclass[conference,twocolumn,10pt,twoside]{IEEEtran} 
\usepackage{setspace,amssymb,graphicx,enumerate,amsfonts,amsmath,color,algorithm,algpseudocode}
\usepackage{epsfig,subfigure,booktabs,amsmath,calc,arydshln,tikz,cite,pdflscape,rotating}
\usepackage[english]{babel}
\usepackage{epstopdf}
\usetikzlibrary{shapes}
\usepackage{calc}
\usepackage{ifthen}
\usepackage{verbatim}
\usepackage{lscape}

\colorlet{myblue}{blue!15}
\colorlet{mygreen}{green!20}
\colorlet{myorange}{orange!20}
\definecolor{mygray}{gray}{0.95}
\DeclareMathOperator{\sign}{sign}

\ifCLASSOPTIONonecolumn
\fi

\IEEEoverridecommandlockouts
\begin{document}
\title{New Algorithms for Maximizing \\ Cellular Wireless Network Energy Efficiency}

\author{\IEEEauthorblockN{Kemal Davaslioglu}
\IEEEauthorblockA{Department of Electrical Engineering\\
University of South Florida\\
Email: kemald@usf.edu}
\and
\IEEEauthorblockN{Cemil Can Coskun and Ender Ayanoglu}
\IEEEauthorblockA{Center for Pervasive Communications and Computing\\
Department of Electrical Engineering and Computer Science, \\
University of California, Irvine\\
Email: \{ccoskun, ayanoglu\}@uci.edu}
}

\maketitle
\begin{abstract} 
In this paper, we aim to maximize the energy efficiency of cellular wireless networks. Specifically, we address the power allocation problem in multi-cell multi-carrier systems. Considering realistic base station power consumption models, we formulate a network-wide energy efficiency maximization problem. Using tools from fractional programming, we cast this problem in the framework of bi-criterion optimization where rate maximization and power minimization are weighted accordingly. Interference pricing mechanism is applied to reduce the inter-cell interference and to achieve a higher network performance. We decompose the main problem into subproblems via dual decomposition. These subproblems are independently solved per sector using limited information exchange between base stations. We first derive our expressions and present algorithms for the single-tier networks. Then, we extend our analysis to two-tier networks where picocell base stations are deployed to improve the network performance and reduce the link distances. Lastly, we extend our framework and include the quality-of-service constraints. We obtain closed-form expressions for the power level updates which are determined by the multi-level water-filling algorithm, or, as it is sometimes called as, the modified water-filling algorithm. Based on our simulation results, we demonstrate that the proposed algorithms can outperform the benchmark approaches in terms of energy efficiency by a factor of 2.7.

\end{abstract}

\section{Introduction}

With the rapid increase in the number of mobile connected devices and continuing demand on higher data rates, there is a need for energy-efficient solutions for wireless networks. Energy efficiency is not going to be achieved through a single solution, but rather will be the result of a cumulative effect of several solutions. These solutions will come in many different flavors such as new enabling technologies (e.g., massive MIMO, device-to-device communications), new architectural changes (e.g., deployment of heterogeneous networks), energy-efficient equipment (e.g., the advances in power amplifiers), protocol changes, etc. \cite{GreenPaper}.  Energy efficiency is important for both the network operators and the end-users. From the perspective of a network operator, energy efficiency means lowering the operational expenses, improving environmental sustainability, and reducing their carbon footprint. From the point of view of an end-user, energy-efficient equipment means longer battery life and mitigation of the energy trap problem, see \cite{GreenPaper}. In this paper, we address these concerns on network power consumption and throughput, and we design algorithms to improve the network energy efficiency. 

\subsection{Related Works}
Related works on energy efficiency maximization problem similar to the one studied in this paper include but are not limited to \cite{GeoffreyLA,GeoffreyDistributed,FettweisChristian12,CCC2014,Palomar08a,Scale09,Palomar13,Palomar14,Buzzi14,Tang15,Geoffrey13}. In \cite{GeoffreyLA} and \cite{GeoffreyDistributed}, the authors study maximizing the energy efficiency of single links consisting of a transmitter and a receiver pair in which their energy efficiency definition includes both the transmit power and power consumed in the circuitry. They demonstrate that the energy efficiency maximization does not always overlap with the throughput maximization. In our paper, we address the same problem but we consider it in a multi-cell scenario which is more complex. The study in \cite{FettweisChristian12} formulates the energy efficiency maximization problem as bi-criterion optimization through the use of fractional programming methods, in which the rate maximization and power minimization problems are proportionally weighted. In our paper, we also pursue the same approach for casting the bi-criterion optimization and employ the Dinkelbach method for root finding. Despite the similarities, our work differs from \cite{FettweisChristian12} in two major points. First, the problem in \cite{FettweisChristian12} is defined for a single-cell energy efficiency maximization, whereas we formulate a multi-cell problem via introducing pricing mechanisms. From a game theoretical point of view, our formulation corresponds to a case where competing players (base stations) cooperate to achieve a higher optimum solution for the sum of their individual profits (energy efficiencies), whereas the one in \cite{FettweisChristian12} corresponds to a non-cooperative scenario where the players compete for resources. This means that the solution proposed in this paper will outperform the one in \cite{FettweisChristian12}, as also demonstrated by our simulation results. Second, in our formulation, we extend the energy efficiency maximizing problem to include other constraints such as the minimum rate, total power constraints, and spectral mask constraints per subcarrier. Our prior work in \cite{CCC2014} addresses the same problem, but it uses constant power allocation across subbands. It employs two variables to characterize the power transmissions per sector, which are to be optimized. In the sequel, we will pursue a different approach and consider allocating different power levels per subcarrier. The fact that we use different power levels on each subcarrier brings an additional gain over the constant power allocation scenario. Therefore, this paper will provide an upper bound for the one in \cite{CCC2014}. Another difference is that, in \cite{CCC2014}, we only implemented power control in macrocell base stations and did not consider it for picocell base stations. In this paper, we will employ power control in both tiers. Also, several recent works have focused on convexifying similar utility maximization problems (see e.g., \cite{Scale09,Palomar08a,Palomar13,Palomar14}) to provide lower bounds on the original objective. As we will see in Section \ref{Section:EEPowerConstraints}, network energy efficiency is defined as the sum of sector energy efficiencies in this paper. One can also define a similar metric such as the generalized energy efficiency which is the ratio of the network sum rate to the sum power dissipated in the network, and this is investigated in \cite{Buzzi14,Tang15}. Lastly, we refer the interested reader to \cite{GreenPaper,Geoffrey13} for comprehensive literature surveys on energy-efficient communications, investigating many energy-efficient resource allocation algorithms for various cognitive radio, cooperative networks, multiple subcarrier, and multiple antenna systems.


Pricing in the resource allocation problem for wireless networks has been widely studied in the literature, see e.g., \cite{Honig06,Berry09Mag,Honig09,Yu07,Furqan13}. Especially, in cellular networks with dense base station deployments, inter-cell interference becomes a limiting factor that needs to be accounted for. Pricing mechanisms offer effective solutions to alleviate interference such that a higher network optimum solution can be achieved. To reduce the interference, the studies in \cite{Honig06,Berry09Mag,Honig09,Yu07} propose to penalize the transmissions based on the interference they create. In order to convey the interference information, called as \emph{interference prices}, limited information exchange between base stations is required. The major difference between our work and the studies in \cite{Honig06,Berry09Mag,Honig09}, which also employ interference pricing, is that we incorporate the interference pricing terms to determine the water-filling levels, whereas those studies have not done so. Thus, we do not need to take any derivatives. In terms of optimality, as the studies in \cite{Honig09,Yu07} also pointed out, the power control problem for the multi-cell networks is a non-convex problem. Due this  non-convex nature, convergence to a global maximum is hard to achieve \cite{Yu07}. The obtained solutions satisfy the Karush-Kuhn-Tucker (KKT) conditions that guarantee convergence to a local maximum. 


\subsection{Contributions}
In this paper, we study the multi-cell multi-carrier network energy efficiency maximization problem. We take into account the transmit power and static power consumption of base stations. The linearized load-dependent power consumption model in \cite{FehskeWC11} is employed. This model considers the contributions of the power amplifier, radio-frequency small-signal transceiver module, baseband receiver unit, power supply, and cooling. Using methods from fractional programming, we reformulate the energy efficiency maximization problem as a bi-criterion optimization problem in which the minimum power and maximum throughput problems are weighted accordingly. We obtain closed-form expressions for the water-filling algorithm. Using dual decomposition and the interference pricing mechanism, we decouple the network-wide energy efficiency problem into subproblems which are solved independently at each sector using limited information exchange. In addition, we incorporate several practical constraints in our formulation. We consider the total transmit power of a base station and the maximum power levels per subcarrier to account for different spectral masks and power amplifier constraints. We also incorporate the minimum rate constraints per user to account for different quality-of-service levels. Since the proposed algorithms employ closed-form expressions for the power updates and do not require any derivatives, their implementation complexities are significantly low compared to the works in \cite{Honig06,Berry09Mag,Honig09,CCC2014}. We evaluate the performance of the proposed algorithm and compare its performance with the ones proposed in \cite{FettweisChristian12} and \cite{CCC2014} and demonstrate that the proposed algorithm outperforms both of these works.

The remainder of this paper is organized as follows. In Section~\ref{Section:EEPowerConstraints}, we formulate the multi-cell energy efficiency maximization problem with power constraints. We derive the corresponding iterative water-filling solution and present the proposed algorithm. We study the same problem for two-tier networks in Section~\ref{Section:EEMaxTwoTier} and extend the preceding framework to include minimum rate constraints in Section~\ref{Section:RateConstraints}. We present the corresponding solution and its implementation steps. Section~\ref{Section:EEWFSimResults} discusses our simulation results, where we evaluate the performance of the proposed algorithm and compare its performance with several benchmarks to quantify the additional gains. Finally, Section~\ref{Section:EEWFQoSOutage} provides the concluding remarks.
\section{Multi-cell Energy Efficiency Maximization Problem with Power Constraints in Single-Tier Networks}
\label{Section:EEPowerConstraints}
In this section, we discuss the energy efficiency maximization problem for the multi-cell multi-carrier systems in a single-tier network. This means that there are only macrocell base stations in the network. We consider three-sector antennas at macrocell base stations. To model the power consumption at a base station, we employ the load-dependent power consumption model proposed in \cite{FehskeWC11}. Our objective is to maximize the sum of sector energy efficiencies in the network subject to the power constraints at each base station. In what follows, we first obtain the power consumption expression in each sector and then define the energy efficiency maximization problem. We denote the power consumed at each macrocell base station sector~$s$ by 
\begin{align} 
\begin{aligned}
P_{\text{Macro,s}}(\textbf{p}_s) = P_{0,s} + \Delta_M \left\lVert \textbf{p}_s \right\lVert_1
\end{aligned}
\end{align}
where $P_{0,s}$ is the power consumption at the minimum non-zero output power of a macrocell sector~$s$ and $\Delta_M$ is the slope of the load-dependent power consumption of macrocell base station sector \cite{FehskeWC11}. The set of subcarriers is denoted by $\mathcal{N}$. The RF output power per subcarrier $n$ at sector~$s$ is represented by $p_{s}^{(n)}$ and the vector $\textbf{p}_s = [p_s^{(1)}, \cdots, p_s^{(N)}]$ is the set of RF output transmit power levels of a macrocell sector~$s$ over $N$ subcarriers. The operator $\left\lVert \cdot \right\lVert_1$ denotes the $\ell_1$-norm. Using the power consumption model, we can formulate the multi-cell multi-carrier network energy efficiency maximization problem for a single-tier network as follows
\begin{align}\max & \hspace{1em} \sum_{s \in \mathcal{S}_m} \left[ \left(\sum_{n \in \mathcal{N}} \Delta_f \log_2 \left(1 + p_{s}^{(n)} \chi_k^{(n)}\right) \right) / P_{\text{Macro,s}}(\textbf{p}_s) \right] \nonumber \\
\text{s.t. } & \hspace{1em} \left\lVert \textbf{p}_s \right\lVert_1 \leq P_{\text{Total,s}} \hspace{.5em} \text{ for all } s \in \mathcal{S}_m \label{Eqn:WFEEFrac} \\
& \hspace{1em} P_{\max,s}^{(n)} \geq p_{s}^{(n)} \geq 0 \text{ for all } n \in \mathcal{N} \text{ and for all } s \in \mathcal{S}_m \nonumber
\end{align}
where $\Delta_f$ is the subcarrier bandwidth and the set of all macrocell sectors is denoted by $\mathcal{S}_m$, the total transmit power of a macrocell base station sector is $P_{\text{Total,s}}$, and the maximum transmit power per subcarrier is denoted by $P_{\max,s}^{(n)}$. Note that the quantity maximized in (\ref{Eqn:WFEEFrac}) has units bits/Joule. The channel-to-interference-plus-noise ratio (CINR) of user~$k$ is 
\begin{align} \chi_k^{(n)} = \frac{g_{k,s}^{(n)}}{\sigma^2 + I_k^{(n)}} = \frac{g_{k,s}^{(n)}}{\left(\sigma^2 + \sum_{s' \neq s, s' \in \mathcal{S}^{(n)}} p_{s'}^{(n)} g_{k,s'}^{(n)}\right)}, \end{align}
where $g_{k,s}^{(n)}$ is the channel gain between user~$k$ and macrocell sector~$s$, and $I_k^{(n)}$ is the interference incurred by user~$k$ on subcarrier~$n$. The set $\mathcal{S}^{(n)}$ is the set of base stations that transmit on subcarrier $n$. Using this notation, $s' \neq s, s' \in \mathcal{S}^{(n)}$ denotes the set of base stations that creates  interference to user $k$ on subcarrier~$n$. In (\ref{Eqn:WFEEFrac}), we maximize the aggregate energy efficiencies of sectors with respect to the total power constraints and per subcarrier power constraints. The first constraint in (\ref{Eqn:WFEEFrac}) is due to the maximum power limitations at the base station, which are defined by the standards. The second constraint in (\ref{Eqn:WFEEFrac}) arises due to the spectral masks \cite{Yu06DSL}. The work in \cite{Dinkelbach67} shows how to relate a fractional program to a parametric program and develops an effective and simple algorithm. In this paper, we will employ the same approach such that the problem in (\ref{Eqn:WFEEFrac}) is translated into the following equivalent form by introducing a new parameter $\lambda_{s}$ per sector 
\begin{align}\max & \hspace{1em} \sum_{s \in \mathcal{S}_m} \left[\sum_{n \in \mathcal{N}} \Delta_f \log_2 \left(1 + p_{s}^{(n)} \chi_k^{(n)}\right) - \lambda_{s} P_{\text{Macro,s}}(\textbf{p}_s) \right] \nonumber \\
\text{s.t. } & \hspace{1em} \left\lVert \textbf{p}_s \right\lVert_1 \leq P_{\text{Total,s}} \hspace{.5em} \text{ for all } s \in \mathcal{S}_m \label{Eqn:WFEE0}\\
& \hspace{1em} P_{\max,s}^{(n)} \geq p_{s}^{(n)} \geq 0 \text{ for all } n \in \mathcal{N} \text{ and for all } s \in \mathcal{S}_m. \nonumber
\end{align}
This type of formulation enables us to obtain closed form expressions. From an optimization perspective, this corresponds to a bi-criterion optimization problem in which both the rate maximization and power consumption minimization are two objectives that we want to jointly solve \cite{OptBookBoyd}. In other words, with this new objective, the rate maximization objective is weighted with one and power consumption minimization objective by $-\lambda_{s}$ at each sector~$s$. When we write the Lagrangian of the problem (\ref{Eqn:WFEE0}), we obtain 
\ifCLASSOPTIONonecolumn 
\label{Eqn:LagrangianWFEE0}
\begin{align} \mathcal{L}(\{\textbf{p}_s\},\boldsymbol{\lambda},\boldsymbol{\mu}) = \sum_{s \in \mathcal{S}_m} \left[\sum_{n \in \mathcal{N}} \Delta_f \log_2 \left(1 + p_{s}^{(n)} \chi_k^{(n)}\right) - \lambda_{s} P_{\text{Macro,s}}(\textbf{p}_s)  + \mu_s \left( P_{\text{Total,s}} - \left\lVert \textbf{p}_s \right\lVert_1 \right) \right], \label{Eqn:LagrangianWFEE0}
\end{align}
\else  
\begin{align}\begin{aligned} & \mathcal{L}(\{\textbf{p}_s\},\boldsymbol{\lambda},\boldsymbol{\mu}) = \sum_{s \in \mathcal{S}_m} \left[\sum_{n \in \mathcal{N}} \Delta_f \log_2 \left(1 + p_{s}^{(n)} \chi_k^{(n)}\right)\right. \\ 
& \hspace{5em}  \left. - \lambda_{s} P_{\text{Macro,s}}(\textbf{p}_s)  + \mu_s \left( P_{\text{Total,s}} - \left\lVert \textbf{p}_s \right\lVert_1 \right) \right]
\end{aligned}\label{Eqn:LagrangianWFEE0}
\end{align}
\fi
where $\boldsymbol{\lambda} = [\lambda_1,\cdots,\lambda_S]$. The vector $\boldsymbol{\mu} = [\mu_1,\cdots,\mu_S]$ denotes the non-negative Lagrange variables associated with the total power at each base station. The transmit powers of all macrocell sectors over all subcarriers are denoted by the set of vectors 
$\{\textbf{p}_s\}  = \{\textbf{p}_1,\cdots,\textbf{p}_S\}$. Optimization theory tells us that the dual function yields lower bounds on the optimal value of the Problem (\ref{Eqn:WFEE0}) \cite{OptBookBoyd}, and it is is given by
\begin{align}
g(\boldsymbol{\lambda},\boldsymbol{\mu}) = \left\{ \begin{array}{ll}
\max\limits_{\{\textbf{p}_s\}} & \mathcal{L}(\{\textbf{p}_s\},\boldsymbol{\lambda},\boldsymbol{\mu}) \\
\text{s.t.}  & P_{\max,s}^{(n)} \geq p_{s}^{(n)} \geq 0  \\ &  \text{for all } n \in \mathcal{N} \text{ and for all } s \in \mathcal{S}_m 
 \end{array} \right.
\end{align}
where the dual function $g(\boldsymbol{\lambda},\boldsymbol{\mu})$ solves for the maximum value of the Lagrangian (\ref{Eqn:LagrangianWFEE0}) for given $\boldsymbol{\lambda}$ and $\boldsymbol{\mu}$. Next, we take the derivative of (\ref{Eqn:LagrangianWFEE0}) with respect to $p_{s}^{(n)}$ and equate the corresponding equation to zero. Then, we obtain
\ifCLASSOPTIONonecolumn
\begin{align}
\frac{\partial \mathcal{L}}{\partial p_{s}^{(n)}} =& \frac{\Delta_f}{\log(2)} \cdot \frac{\chi_k^{(n)}}{1+p_{s}^{(n)} \chi_k^{(n)}} - \frac{\Delta_f}{\log(2)} \sum\limits_{j \neq k, j \in \mathcal{K}^{(n)}}  \pi_{k,j}^{(n)} - \lambda_{s} \cdot \Delta_M  - \mu_s = 0,
\end{align}
\else
\begin{align}
\frac{\partial \mathcal{L}}{\partial p_{s}^{(n)}} =& \frac{\Delta_f}{\log(2)} \cdot \frac{\chi_k^{(n)}}{1+p_{s}^{(n)} \chi_k^{(n)}} - \frac{\Delta_f}{\log(2)} \sum\limits_{j \neq k, j \in \mathcal{K}^{(n)}}  \pi_{k,j}^{(n)} \nonumber
\\ & - \lambda_{s} \cdot \Delta_M  - \mu_s  = 0,
\end{align}
\fi
where the interference pricing terms are expressed as
\begin{align}
\pi_{k,j}^{(n)} = \frac{\gamma_j^{(n)}}{\gamma_j^{(n)} + 1} \cdot \frac{g_{j,s}^{(n)}}{I_j^{(n)} + \sigma^2},
\end{align}
and where $\gamma_{j}^{(n)}$ is the signal-to-noise-ratio of user~$j$ on subcarrier~$n$. The set of users assigned to subcarrier~$n$ is given by $\mathcal{K}^{(n)}$.
Then, the set $j \neq k, j \in \mathcal{K}^{(n)}$ denotes the set of users that sector~$s$ interferes on subcarrier~$n$ while transmitting to its associated user~$k$. When we fix the interference prices and the power levels of base stations except for sector~$s$, and rearrange terms, we have the following closed-form expression for the transmit power allocated to user~$k$ of sector~$s$ on subcarrier~$n$ 
\ifCLASSOPTIONonecolumn
\begin{align}\label{Eqn:EEWFPricing}
p_{s}^{(n)}  = \left[\frac{1}{ \frac{\log(2)}{\Delta_f} \cdot \left(\lambda_{s} \cdot \Delta_M   + \mu_s \right) + \sum\limits_{j \neq k, j \in \mathcal{K}^{(n)}}  \pi_{k,j}^{(n)}} - \frac{1}{\chi_{k}^{(n)}}  \right]_0^{P_{\max,s}^{(n)}},
\end{align} 
\else 
\begin{align}\label{Eqn:EEWFPricing}
\resizebox{\linewidth}{!}{$p_{s}^{(n)}  = \left[\frac{1}{ \frac{\log(2)}{\Delta_f} \cdot \left(\lambda_{s} \cdot \Delta_M   + \mu_s \right) + \sum\limits_{j \neq k, j \in \mathcal{K}^{(n)}}  \pi_{k,j}^{(n)}} - \frac{1}{\chi_{k}^{(n)}}  \right]_0^{P_{\max,s}^{(n)}}$},
\end{align} 
\fi
where $[x]_0^{P_{\max,s}}$ denotes that $x$ is lower bounded by $0$ and upper bounded by $P_{\max,s}^{(n)}$. Equation (\ref{Eqn:EEWFPricing}) suggests that whenever the transmissions of a sector create high interference to the users in neighboring cells, the water-filling levels are reduced, and the corresponding transmissions are decreased. The closed-form expression in (\ref{Eqn:EEWFPricing}) closely depends on the value of $\mu_s$. It is straightforward to show that $\left\lVert \textbf{p}_s(\mu_s) \right\lVert_1   \leq P_{\text{Total,s}}$, where the transmit power at subcarrier~$n$ is a function of $\mu_s$. As the value of $\mu_s$ increases, the aggregate transmit power monotonically decreases. We employ a one-dimensional search such as the bisection algorithm under the heading Algorithm~\ref{Algorithm:Bisection} to find the optimal $\mu_s^\ast$ that satisfies the sum power constraints. In Algorithm~\ref{Algorithm:Bisection}, we first determine the search domain for the bisection algorithm, where the lower bound $\mu_{s,l}$ is set to zero, while the upper bound $\mu_{s,u}$ is increased to the powers of two until the aggregate transmit power is below $P_{\text{Total,s}}$. When $\mu_{s,u}$ is found, the algorithm proceeds to the classical binary search procedure. The loop terminates when the difference between the upper and lower values is less than the threshold. Finally, if the sum of transmit powers is less than the total power constraint, then $\mu_{s,mid}$ needs to be set to zero, which comes from the complementary slackness condition \cite{Yu07,OptBookBoyd}.

\begin{algorithm}[t!]
  \caption{Bisection Method for the Iterative Water-Filling Algorithm}\label{Algorithm:Bisection}
      \begin{algorithmic}[1]
	\State Let $\epsilon$ denote the tolerance and $l_{\max}$ be the maximum number of iterations. Initialize $\mu_{s,l} = 0$ and $\mu_{s,u} = 1$
	\State Calculate $p_{s}^{(n)}(\mu_{s,u})$. 
	\While {$\left\lVert \textbf{p}_s(\mu_{s,u}) \right\lVert_1  > P_{\text{Total,s}}$}
	\State $\mu_{s,u} = 2 \times \mu_{s,u}$
	\EndWhile{}	
	\While {$|\mu_{s,u} - \mu_{s,l}| > \epsilon$}
	\State $\mu_{s,mid} = (\mu_{s,l} + \mu_{s,u})/2$
	\State Calculate $p_s^{(n)}(\mu_{s,mid})$ using (\ref{Eqn:EEWFPricing})
	\If{$\sign\left( \left\lVert \textbf{p}_s(\mu_{s,mid}) \right\lVert_1 - P_{\text{Total,s}}\right) = \sign\left( \left\lVert \textbf{p}_s(\mu_{s,l}) \right\lVert_1  - P_{\text{Total,s}}\right)$}	
	\State $\mu_{s,l} = \mu_{s,mid}$
	\Else
	\State $\mu_{s,u} = \mu_{s,mid}$
	\EndIf
	 \EndWhile{}	
	 \If{ $ \sign\left( \left\lVert \textbf{p}_s(\mu_{s,mid}) \right\lVert_1 \right)  < P_{\text{Total,s}}$}
	 \State $\mu_{s,mid} = 0$
	 \EndIf 
  \end{algorithmic}
\end{algorithm}


Let the optimal cut-off value in the water-filling solution be defined as
\begin{align} \Omega_{EE,P}^{\ast (n)} =   \frac{\log(2)}{\Delta_f} \cdot \left(\lambda_{s} \cdot \Delta_M   + \mu_s \right) + \sum\limits_{j \neq k, j \in \mathcal{K}^{(n)}}  \pi_{k,j}^{(n)},\label{Eqn:OmegaEE}\end{align}
where the initials EE and P stand for energy efficiency maximization and pricing, respectively. In the water-filling solution, this cut-off value can be interpreted as the threshold that determines if the subcarrier is used or not. Any subcarrier $n$ with the CINR, $\chi_k^{(n)}$, is not used if its magnitude is below the optimal cut-off value $\Omega_{EE,P}^{\ast (n)}$. Mathematically, we can express this condition as 
\begin{align} p_{s}^{(n)} > 0 & \text{ if }\Omega_{EE,P}^{\ast (n)} <\chi_{k}^{(n)} \text{ and }  p_{s}^{(n)} = 0 \text{ if }\Omega_{EE,P}^{\ast (n)} \geq \chi_{k}^{(n)}. \end{align}
Notice that the cut-off value depends both on frequency-dependent and frequency-independent terms. Frequency-dependent terms come from the interference pricing values, denoted by $\pi_{k,j}^{(n)}$, while the frequency-independent terms are system related parameters such as $\Delta_M$, $\Delta_f$, and $\mu_s$. In an interference-dominated region, the water-filling levels are adjusted based on the interference pricing terms.

\begin{figure}[t!]
\centering
		\ifCLASSOPTIONonecolumn
			\includegraphics[width=0.6\columnwidth]{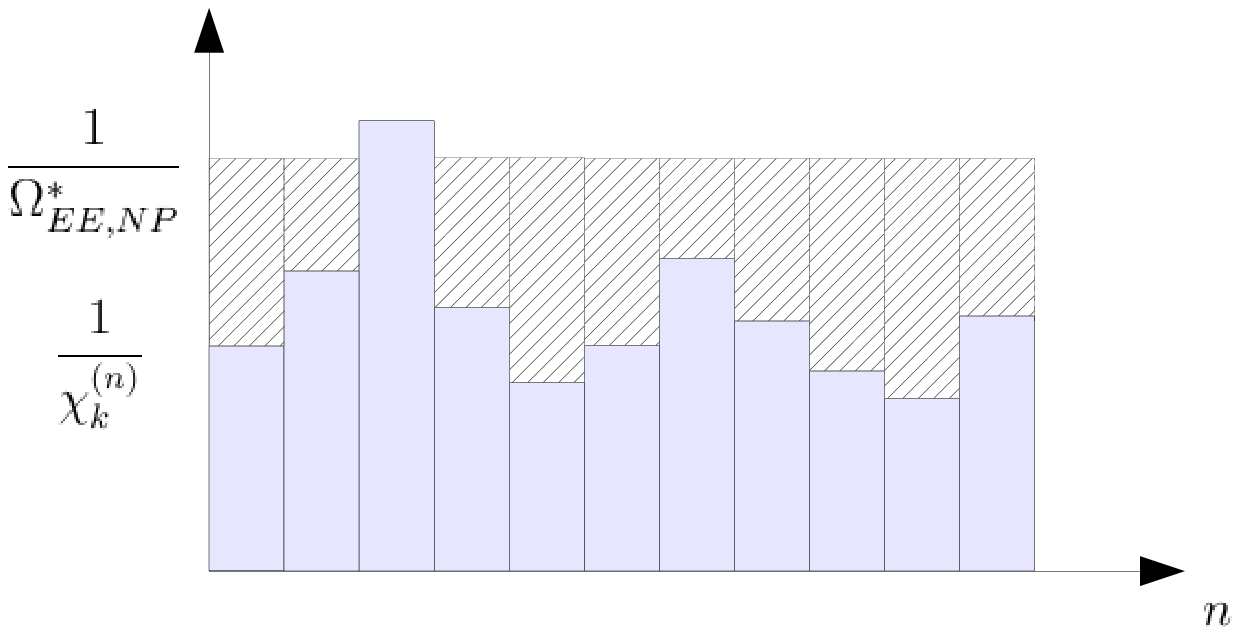}
		\else
			\includegraphics[width=0.85\columnwidth]{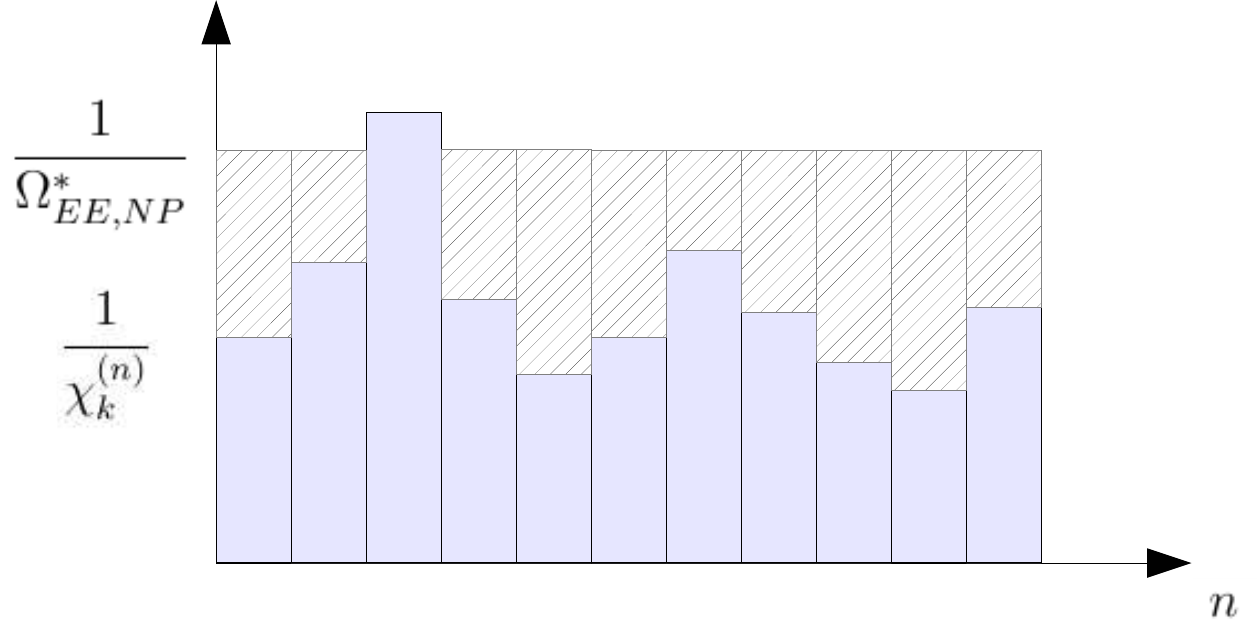}
		\fi		
		\caption{Single level water-filling solution for energy efficiency maximization without pricing.}  
		\label{Figure:SingleLevelWF}
\end{figure}

\begin{figure}[th!]
\centering
		\ifCLASSOPTIONonecolumn
			\includegraphics[width=0.6\columnwidth]{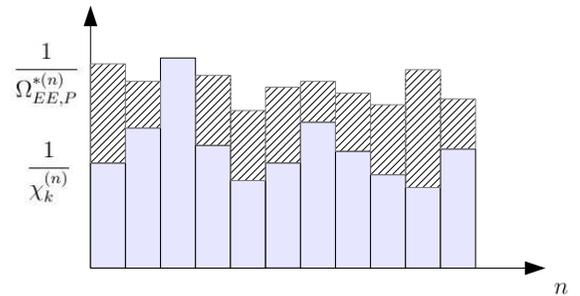}
		\else
					\includegraphics[width=0.85\columnwidth]{MultiLevelWFShaded.eps}
		\fi	
		\caption{Multi-level water-filling solution for energy efficiency maximization where the pricing terms determine the water filling level on each subcarrier.}  
		\label{Figure:MultiLevelWF}
\end{figure}

The closed-form expression in (\ref{Eqn:EEWFPricing}) corresponds to the solution for the energy-efficient maximization problem with pricing. For the case without pricing, the closed-form expression reduces to
\begin{align}\label{Eqn:EEWFNoPricing}
p_{s}^{(n)}  = \left[\frac{1}{ \frac{\log(2)}{\Delta_f} \cdot \left(\lambda_{s} \cdot \Delta_M   + \mu_s \right) } - \frac{1}{\chi_{k}^{(n)}}  \right]_0^{P_{\max,s}^{(n)}}.
\end{align}
Similarly, the cut-off value for the case without pricing is given by 
\begin{align} \Omega_{EE,NP}^\ast  = \frac{\log(2)}{\Delta_f} \cdot \left(\lambda_{s} \cdot \Delta_M   + \mu_s \right), \end{align}
where the initials NP stand for no pricing case. Note that, in the case without pricing, the cut-off value is constant for all subcarriers and it has no frequency dependency since interference pricing terms are omitted in the solution. In Figures~\ref{Figure:SingleLevelWF} and~\ref{Figure:MultiLevelWF}, we illustrate examples of water-filling energy efficiency maximization solutions without and with pricing, respectively. In Fig.~\ref{Figure:SingleLevelWF}, we observe that the optimal water-filling level is constant throughout the subcarriers, and thus, there is a single level for water-filling. When we incorporate interference pricing, we observe that there are multiple levels for water-filling level on each subcarrier. When the created interference is high on particular subcarriers, i.e., higher interference prices, the water-filling levels are lowered.

		
Another way of relating the problems in (\ref{Eqn:WFEEFrac}) and (\ref{Eqn:WFEE0}) is as follows. Let $q_s^\ast$ and $\lambda_{s}^\ast$ denote the respective solutions for these two problems in the same order as before. For each sector, consider the following function
\begin{align} F_s(\lambda_{s})  =  \max_{\textbf{p}_s}\hspace{-0.01in} \sum\limits_{n \in \mathcal{N}} \hspace{-0.03in} \Delta_f \log_2 \left(1 + p_{s}^{(n)} \chi_k^{(n)}\right) - \lambda_{s}  P_{\text{Macro,s}}(\textbf{p}_s)
\end{align}
where the vector $\textbf{p}_s$ satisfies the feasibility conditions, i.e., $\{\textbf{p}_s \in \mathcal{P} | \sum_{n \in \mathcal{N}} p_{s}^{(n)} \leq P_{\text{Total,s}};$ $P_{\max,s}^{(n)} \geq p_{s}^{(n)} \geq 0 \}$ for all $n \in \mathcal{N}$ and $s \in \mathcal{S}_m$.
Then, the following statements are true \cite{FettweisChristian12,Dinkelbach67}:
\begin{align}\begin{aligned} 
F_s(\lambda_{s})  > 0, & \quad \text{if } \lambda_s < q_s^\ast \\
F_s(\lambda_{s})  = 0, & \quad \text{if }\lambda_s = q_s^\ast \\ 
F_s(\lambda_{s})  < 0, & \quad \text{if } \lambda_s > q_s^\ast.
\end{aligned}\end{align}
Hence, solving problem (\ref{Eqn:WFEEFrac}) is equivalent to finding the roots of $F_s(\lambda_{s})$, and the corresponding optimal condition is 
\ifCLASSOPTIONonecolumn
\begin{align}
F_s(\lambda_{s}^\ast) = \max_{\textbf{p}} & \left( \sum_{n \in \mathcal{N}} \Delta_f \log_2 \left(1 + p_{s}^{(n)} \chi_k^{(n)}\right) - \lambda_{s}^\ast P_{\text{Macro,s}(\textbf{p}_s) }\right) = 0.
\end{align}
\else
\begin{align}\resizebox{\linewidth}{!}{$F_s(\lambda_{s}^\ast) = \max\limits_{\textbf{p}}\left( \sum_{n \in \mathcal{N}} \Delta_f \log_2 \left(1 + p_{s}^{(n)} \chi_k^{(n)}\right) - \lambda_{s}^\ast P_{\text{Macro,s}}(\textbf{p}_s) \right) = 0.$}
\end{align}
\fi
\ifCLASSOPTIONonecolumn
\begin{algorithm}[t!]
\caption{Iterative Water-Filling Algorithm with Pricing for Network Energy Efficiency Maximization}\label{Algorithm:EEWFAlgPricing}
      \begin{algorithmic}[1]
	\State Initialize transmit power levels and interference prices and set  $t = 0$. Solve the following at each sector~$s$	
 	\While{$|F_s(\lambda_{s})| > \epsilon$  and $l < l_{\max}$} \label{Step:EEMax}
    \State $\lambda_{s}(l) = \left(\sum_{n \in \mathcal{N}} \Delta_f \log_2 \left(1 + p_{s}^{(n)} \chi_k^{(n)}\right)\right)/P_{\text{Macro,s}}(\textbf{p}_s) $
	\State Obtain $\mu_{s}$ using the bisection method \label{Step:EEMaxBisection}
    \State For all $n \in \mathcal{N}$, solve for $p_{\text{Next}}^{(n)}$ using  (\ref{Eqn:EEWFPricing})
    \State Calculate $F_s(\lambda_{s})$    
	\State Update $l = l + 1$
	\EndWhile{}
	\State Update the power levels using
    \begin{align} p_{s}^{(n)}(t + 1) = (1-\delta) \cdot p_{s}^{(n)}(t) + \delta \cdot p_{\text{Next}}^{(n)}
    \end{align}
    \State Distribute the interference prices, $\{ \pi_{k,j}^{(n)}\}$, among base stations
    \State Go to Step~\ref{Step:EEMax} and repeat for $t=t+1$
\end{algorithmic}
\end{algorithm}
\else
\begin{algorithm}[t!]
\caption{Iterative Water-Filling Algorithm with Pricing for Network Energy Efficiency Maximization}\label{Algorithm:EEWFAlgPricing}
      \begin{algorithmic}[1]
      	\State Initialize transmit power levels and interference prices and set  $t = 0$. Solve the following at each sector~$s$	
 	\While{$|F_s(\lambda_{s})| > \epsilon$  and $l < l_{\max}$} \label{Step:EEMax}
    \State \resizebox{0.9\linewidth}{!}{$\lambda_{s}(l) = \left(\sum_{n \in \mathcal{N}} \Delta_f \log_2 \left(1 + p_{s}^{(n)} \chi_k^{(n)}\right)\right)/P_{\text{Macro,s}}(\textbf{p}_s) $}
	\State Obtain $\mu_{s}$ using the bisection method \label{Step:EEMaxBisection}
    \State For all $n \in \mathcal{N}$, solve for $p_{\text{Next}}^{(n)}$ using using  (\ref{Eqn:EEWFPricing})
    \State Calculate $F_s(\lambda_{s})$    
	\State Update $l = l + 1$
	\EndWhile{}
	\State Update the power levels using
    \begin{align} p_{s}^{(n)}(t + 1) = (1-\delta) \cdot p_{s}^{(n)}(t) + \delta \cdot p_{\text{Next}}^{(n)}
    \end{align}
    \State Distribute the interference prices, $\{ \pi_{k,j}^{(n)}\}$, among base stations
    \State Go to Step~\ref{Step:EEMax} and repeat for $t=t+1$
\end{algorithmic}
\end{algorithm}
\fi


We summarize the iterative energy-efficient water-filling algorithm with pricing under the heading Algorithm~\ref{Algorithm:EEWFAlgPricing} in which the variables $\textbf{p}$, $\lambda_s$, and $\mu_s$ are iteratively updated.
We use the Dinkelbach method to update $\lambda_s$ at each sector, which is an application of the classical Newton's method for root finding \cite{HandbookCO}. This method has the following iterations 
\begin{align}\begin{aligned}\lambda_s(l) =& \lambda_s(l-1) - \frac{F_s\left(\lambda_s(l-1)\right)}{F_s'\left(\lambda_s(l-1)\right)} \\ =& \frac{\sum\limits_{n \in \mathcal{N}} \Delta_f \log_2 \left(1 + p_{s}^{(n)} \chi_k^{(n)}\right)}{P_{\text{Macro,s}}(\textbf{p}_s) }, \end{aligned}\end{align}
where $F_s'(\lambda_s)$ denotes the derivative of $F_s(\lambda_s)$ with respect to $\lambda_s$. 
Next, we use $\lambda_s(l)$ to determine the power levels, $p_s^{(n)}$, and use Algorithm~\ref{Algorithm:Bisection} to find the optimal $\mu_s$ value. In order to avoid rapid fluctuations which may cause unstability in the system, we use the following technique where the power control parameters are updated as
\begin{align} p_{s}^{(n)}(t+1)  =  (1 - \delta)  \cdot p_{s}^{(n)}(t) + \delta \cdot p_k^{\text{Next}}, \end{align}
where $\delta$ satisfies $\delta(t = 0) = 1$, $\delta(t) \in (0,1)$ for $t > 0$, and $\sum_{t = 0}^{\infty} \delta(t) = \infty$ \cite[p.~286]{GameTheoryBaser}.
In general, $\delta(t)$ is chosen as $\delta(t) = t/(2t+1)$ for $t > 0$. As the limit goes to infinity, $\delta (t)$ approaches $1/2$. This iterative update method is called as the Mann iterative method \cite{GameTheoryBaser}. It is important to note that, without this iterative method, the power level updates may yield large oscillations and sometimes may not converge. It is straightforward to derive a similar algorithm for the network throughput maximization. Although we omit laying out algorithmic steps explicitly due to space considerations, in Section~\ref{Section:EEWFSimResults}, we evaluate its performance and compare it to the network energy efficiency maximization problem.

\section{Multi-cell Energy Efficiency Maximization Problem with Power Constraints in Two-Tier Networks}\label{Section:EEMaxTwoTier}
Next, we consider network energy efficiency maximization for two-tier networks. We consider picocell deployments underlying the macrocell tier and our objective is to maximize the sum of energy efficiencies of all sectors. First, we need to express the total power consumed in a sector as 
\begin{align} \psi_s(\textbf{p}_s, \left\{ \textbf{p}_p \right\}) = P_{\text{Macro,s}}(\textbf{p}_s)  + \sum_{p \in \mathcal{S}_{\text{P,s}}} P_{\text{Pico,p}}(\textbf{p}_p) ,  \end{align}
where $P_{\text{Pico,p}}(\textbf{p}_p)$ denotes total power consumption of a picocell base station~$p$ and the RF output transmit power of picocell base station $p$ over $N$ subcarriers is represented by the vector $\textbf{p}_p$. The set of picocell power levels in sector $s$ is denoted by $\left\{ \textbf{p}_p \right\}$. The set $\mathcal{S}_{\text{P,s}}$ is the set of picocell base stations in sector~$s$. The power consumption at a picocell base station is given by
\begin{align} P_{\text{Pico,p}}(\textbf{p}_p) =& P_{0,p} + \Delta_P \left\lVert \textbf{p}_p \right\lVert_1, \end{align}
where $P_{0,p}$ and $\Delta_P$ are the power consumption at the minimum non-zero output power and the slope of the power consumption of a picocell base station~$p$, respectively. We can now formulate the multi-cell energy efficiency maximization for two-tier networks as follows:
\ifCLASSOPTIONonecolumn
\begin{align}\label{Eqn:WFEEFracTwoTier} \begin{aligned} \max & \hspace{1em} \sum_{s \in \mathcal{S}_m} \left[ \sum_{n \in \mathcal{N}} \Delta_f \log_2\left(1 + p_s^{(n)} \chi_k^{(n)}\right) + \sum_{p \in \mathcal{S}_{P,s}} \sum_{n \in \mathcal{N}}\Delta_f \log_2\left(1+ p_p^{(n)} \chi_k^{(n)}\right) - \lambda_s \psi_s(\textbf{p}_s,\left\{ \textbf{p}_p \right\}) \right]\\
\text{s.t. } & \hspace{1em} \left\lVert \textbf{p}_s \right\lVert_1 \leq P_{\text{Total,s}} \hspace{.5em} \text{ for all } s \in \mathcal{S}_m \\
& \hspace{1em} \left\lVert \textbf{p}_p \right\lVert_1 \leq P_{\text{Total,p}} \hspace{.5em} \text{ for all } s \in \mathcal{S}_{P,s} \\
& \hspace{1em} P_{\max,s}^{(n)} \geq p_{s}^{(n)} \geq 0 \text{ for all } n \in \mathcal{N} \text{ and for all } s \in \mathcal{S}_m \\
& \hspace{1em} P_{\max,p}^{(n)} \geq p_{p}^{(n)} \geq 0 \text{ for all } n \in \mathcal{N} \text{ and for all } s \in \mathcal{S}_{P,s},
\end{aligned}
\end{align}
\else
\begin{align} \max & \hspace{1em} \sum_{s \in \mathcal{S}_m} \left[ \sum_{n \in \mathcal{N}} \Delta_f \log_2\left(1 + p_s^{(n)} \chi_k^{(n)}\right) \right. \nonumber \\ 
&  \hspace{-1em}+ \left. \sum_{p \in \mathcal{S}_{P,s}} \sum_{n \in \mathcal{N}}\Delta_f \log_2\left(1+ p_p^{(n)} \chi_k^{(n)}\right) - \lambda_s \psi_s(\textbf{p}_s,\left\{ \textbf{p}_p \right\}) \right] \nonumber\\
\text{s.t. } & \hspace{1em} \left\lVert \textbf{p}_s \right\lVert_1 \leq P_{\text{Total,s}} \hspace{.5em} \text{ for all } s \in \mathcal{S}_m  \label{Eqn:WFEEFracTwoTier}\\
& \hspace{1em} \left\lVert \textbf{p}_p \right\lVert_1 \leq P_{\text{Total,p}} \hspace{.5em} \text{ for all } s \in \mathcal{S}_{P,s} \nonumber\\
& \hspace{1em} P_{\max,s}^{(n)} \geq p_{s}^{(n)} \geq 0 \text{ for all } n \in \mathcal{N} \text{ and for all } s \in \mathcal{S}_m \nonumber\\
& \hspace{1em} P_{\max,p}^{(n)} \geq p_{p}^{(n)} \geq 0 \text{ for all } n \in \mathcal{N} \text{ and for all } s \in \mathcal{S}_{P,s}, \nonumber
\end{align}
\fi
where $P_{\text{Total,p}}$ and $P_{\max,p}^{(n)}$ are the total transmit power of a picocell base station~$p$ and maximum transmit power of $p$ on subcarrier~$n$, respectively. When we apply Lagrangian relaxation, take the derivative with respect to $p_p^{(n)}$, equate to zero, and rearrange the terms, we obtain the following closed-form expression of the iterative power updates for picocell base stations, which are given by
\ifCLASSOPTIONonecolumn
\begin{align} 
p_{p}^{(n)}  = \left[\frac{1}{ \frac{\log(2)}{\Delta_f} \cdot \left(\lambda_{s} \cdot \Delta_P + \mu_p \right) + \sum\limits_{j \neq k, j \in \mathcal{K}^{(n)}}  \pi_{k,j}^{(n)}} - \frac{1}{\chi_{k}^{(n)}}  \right]_0^{P_{\max,p}^{(n)}},
\end{align}
\else
\begin{align} 
\resizebox{\linewidth}{!}{$p_{p}^{(n)}  = \left[\frac{1}{ \frac{\log(2)}{\Delta_f} \cdot \left(\lambda_{s} \cdot \Delta_P + \mu_p \right) + \sum\limits_{j \neq k, j \in \mathcal{K}^{(n)}}  \pi_{k,j}^{(n)}} - \frac{1}{\chi_{k}^{(n)}}  \right]_0^{P_{\max,p}^{(n)}},$}
\end{align}
\fi
where $\mu_p$ is the dual variable associated with the total power constraint of a picocell base station~$p$. Note that the expression for $p_s^{(n)}$ remains the same as in (\ref{Eqn:EEWFPricing}).

\section{Multi-Cell Energy Efficiency Maximization with Rate and Power Constraints}
\label{Section:RateConstraints}
We now extend the preceding framework and include the minimum rate constraints per user.  The multi-cell multi-carrier network energy efficiency maximization with power constraints and minimum rate constraints can be formulated as 
\ifCLASSOPTIONonecolumn
\begin{align}\label{Eqn:WFEEFracRateConstraints} \begin{aligned} \max & \hspace{1em} \sum_{s \in \mathcal{S}_m} \left[ \sum_{n \in \mathcal{N}} \Delta_f \log_2\left(1 + p_s^{(n)} \chi_k^{(n)}\right) + \sum_{p \in \mathcal{S}_{P,s}} \sum_{n \in \mathcal{N}}\Delta_f \log_2\left(1+ p_p^{(n)} \chi_k^{(n)}\right) - \lambda_s \psi_s(\textbf{p}_s,\left\{ \textbf{p}_p \right\}) \right]\\
\text{s.t. } & \hspace{-1em} \sum_{n \in \mathcal{N}_k} r_k^{(n)} \geq R_{\min,k}, \text{ for all } k \in \mathcal{K} \\
& \hspace{1em} \left\lVert \textbf{p}_s \right\lVert_1 \leq P_{\text{Total,s}} \hspace{.5em} \text{ for all } s \in \mathcal{S}_m \\
& \hspace{1em} \left\lVert \textbf{p}_p \right\lVert_1 \leq P_{\text{Total,p}} \hspace{.5em} \text{ for all } p \in \mathcal{S}_{P,s} \\
& \hspace{1em} P_{\max,s}^{(n)} \geq p_{s}^{(n)} \geq 0 \text{ for all } n \in \mathcal{N} \text{ and for all } s \in \mathcal{S}_m \\
& \hspace{1em} P_{\max,p}^{(n)} \geq p_{p}^{(n)} \geq 0 \text{ for all } n \in \mathcal{N} \text{ and for all } p\in \mathcal{S}_{P,s},
\end{aligned}
\end{align}
\else
\begin{align}\max & \hspace{1em} \sum_{s \in \mathcal{S}_m} \left[ \sum_{n \in \mathcal{N}} \Delta_f \log_2\left(1 + p_s^{(n)} \chi_k^{(n)}\right) \right. \nonumber\\ & \left. + \sum_{p \in \mathcal{S}_{P,s}} \sum_{n \in \mathcal{N}}\Delta_f \log_2\left(1+ p_p^{(n)} \chi_k^{(n)}\right) - \lambda_s \psi_s(\textbf{p}_s,\left\{ \textbf{p}_p \right\}) \right] \nonumber\\
\text{s.t. } & \hspace{1em} \sum_{n \in \mathcal{N}_k} r_k^{(n)} \geq R_{\min,k}, \text{ for all } k \in \mathcal{K} \label{Eqn:WFEEFracRateConstraints}\\
& \hspace{1em} \left\lVert \textbf{p}_s \right\lVert_1 \leq P_{\text{Total,s}} \hspace{.5em} \text{ for all } s \in \mathcal{S}_m \nonumber\\
& \hspace{1em} \left\lVert \textbf{p}_p \right\lVert_1 \leq P_{\text{Total,p}} \hspace{.5em} \text{ for all } p \in \mathcal{S}_{P,s} \nonumber\\
& \hspace{1em} P_{\max,s}^{(n)} \geq p_{s}^{(n)} \geq 0 \text{ for all } n \in \mathcal{N} \text{ and for all } s \in \mathcal{S}_m \nonumber\\
& \hspace{1em} P_{\max,p}^{(n)} \geq p_{p}^{(n)} \geq 0 \text{ for all } n \in \mathcal{N} \text{ and for all } p\in \mathcal{S}_{P,s}, \nonumber
\end{align}
\fi
where $R_{\min,k}$ denotes the minimum rate requirement of user $k$. As we consider multi-carrier systems, the aggregate throughput of subcarriers assigned to a user defines its rate. First constraint in (\ref{Eqn:WFEEFracRateConstraints}) ensures that a user gets at least its minimum rate requirement. Similar to our previous discussion, we introduce $\lambda_{s}$ per sector and the corresponding Lagrangian of the problem (\ref{Eqn:WFEEFracRateConstraints}) can be written as
\ifCLASSOPTIONonecolumn
\begin{align}\label{Eqn:LagrangianWFEE0RateConstraints} \begin{aligned}\mathcal{L}(\textbf{p},\boldsymbol{\lambda},\boldsymbol{\tau},\boldsymbol{\mu}) = &
\sum_{s \in \mathcal{S}_m} \left[ \sum_{n \in \mathcal{N}} \Delta_f \log_2\left(1 + p_s^{(n)} \chi_k^{(n)}\right) + \sum_{p \in \mathcal{S}_{P,s}} \sum_{n \in \mathcal{N}}\Delta_f \log_2\left(1+ p_p^{(n)} \chi_k^{(n)}\right)  \right. \\
& - \lambda_s \psi_s(\textbf{p}_s,\left\{ \textbf{p}_p \right\}) + \sum_{k \in \mathcal{K}_s} \tau_k \left( \sum_{n \in \mathcal{N}_k} r_k^{(n)} - R_{\min,k}\right) + \mu_s \left(P_{\text{Total,s}} - \left\lVert \textbf{p}_s \right\lVert_1 \right) \\
& \hspace{.5in}+ \left. \sum_{p \in \mathcal{S}_{p,s}} \mu_p \left(P_{\text{Total,p}} - \left\lVert \textbf{p}_p \right\lVert_1 \right) \right],
\end{aligned} \end{align}
\else
\begin{align} & \mathcal{L}(\textbf{p},\boldsymbol{\lambda},\boldsymbol{\tau},\boldsymbol{\mu}) = 
\sum_{s \in \mathcal{S}_m} \left[ \sum_{n \in \mathcal{N}} \Delta_f \log_2\left(1 + p_s^{(n)} \chi_k^{(n)}\right)  \right. \nonumber\\
& \hspace{1em} + \sum_{p \in \mathcal{S}_{P,s}} \sum_{n \in \mathcal{N}}\Delta_f \log_2\left(1+ p_p^{(n)} \chi_k^{(n)}\right) - \lambda_s \psi_s(\textbf{p}_s,\left\{ \textbf{p}_p \right\})  \nonumber \\
&  + \sum_{k \in \mathcal{K}_s} \tau_k \left( \sum_{n \in \mathcal{N}_k} r_k^{(n)} - R_{\min,k}\right) + \mu_s \left(P_{\text{Total,s}} - \left\lVert \textbf{p}_s \right\lVert_1 \right) \nonumber\\
& \hspace{1em} + \left. \sum_{p \in \mathcal{S}_{p,s}} \mu_p \left(P_{\text{Total,p}} - \left\lVert \textbf{p}_p \right\lVert_1\right) \right],
\label{Eqn:LagrangianWFEE0RateConstraints}
\end{align}
\fi
where $\boldsymbol{\tau} = [\tau_1,\cdots,\tau_K]$ denotes the vector of Lagrange multipliers associated with the minimum rate constraints and $K$ is the total number of users in the system. The throughput of a user is the sum throughput of subcarriers assigned to this user. For a macrocell-associated user, $r_k^{(n)} = \Delta_f \log_2\left(1 + p_{s}^{(n)} \chi_k^{(n)}\right)$, whereas for a picocell-associated user it is defined as $r_k^{(n)} = \Delta_f \log_2\left(1 + p_{p}^{(n)} \chi_k^{(n)}\right)$. To obtain the closed-form expressions for the macrocell base station power updates, we  take the derivative of (\ref{Eqn:LagrangianWFEE0RateConstraints}) with respect to $p_{s}^{(n)}$, equate it to zero, rearrange terms, and obtain the following closed-form expression for the power levels on each subcarrier
\ifCLASSOPTIONonecolumn 
\begin{align}\label{Eqn:EEWFPricingRateConstraintMacro}
p_{s}^{(n)}  = \left[\frac{\left(1+\tau_k \right)}{\frac{\log(2)}{\Delta_f} \cdot \left(\lambda_{s} \cdot \Delta_M   + \mu_s \right) + \sum\limits_{j \neq k, j \in \mathcal{K}^{(n)}} \left(1 + \tau_j \right)  \pi_{k,j}^{(n)}} - \frac{1}{\chi_{k}^{(n)}}  \right]_0^{P_{\max,s}^{(n)}}.
\end{align}
\else
\begin{align}\label{Eqn:EEWFPricingRateConstraintMacro}
\resizebox{\linewidth}{!}{$p_{s}^{(n)}  = \left[\frac{\left(1+\tau_k \right)}{\log(2)/\Delta_f \cdot \left(\lambda_{s} \cdot \Delta_M   + \mu_s \right) + \sum\limits_{j \neq k, j \in \mathcal{K}^{(n)}} \left(1 + \tau_j \right)  \pi_{k,j}^{(n)}} - \frac{1}{\chi_{k}^{(n)}}  \right]_0^{P_{\max,s}^{(n)}}.$}
\end{align}
\fi
The picocell base station power updates are given as
\ifCLASSOPTIONonecolumn
\begin{align}\label{Eqn:EEWFPricingRateConstraintPico}
p_{p}^{(n)}  = \left[\frac{\left(1+\tau_k \right)}{\frac{\log(2)}{\Delta_f} \cdot \left(\lambda_{s} \cdot \Delta_P  + \mu_s \right) + \sum\limits_{j \neq k, j \in \mathcal{K}^{(n)}} \left(1 + \tau_j \right)  \pi_{k,j}^{(n)}} - \frac{1}{\chi_{k}^{(n)}}  \right]_0^{P_{\max,p}^{(n)}}.
\end{align}
\else
\begin{align}\label{Eqn:EEWFPricingRateConstraintPico}
\resizebox{\linewidth}{!}{$p_{p}^{(n)}  = \left[\frac{\left(1+\tau_k \right)}{\log(2)/\Delta_f \cdot \left(\lambda_{s} \cdot \Delta_P  + \mu_s \right) + \sum\limits_{j \neq k, j \in \mathcal{K}^{(n)}} \left(1 + \tau_j \right)  \pi_{k,j}^{(n)}} - \frac{1}{\chi_{k}^{(n)}}  \right]_0^{P_{\max,p}^{(n)}}.$}
\end{align}
\fi
When the user minimum rate constraint is satisfied, its corresponding Lagrangian multiplier is zero. In that case,  (\ref{Eqn:EEWFPricingRateConstraintMacro}) reduces to  (\ref{Eqn:EEWFPricing}). In addition, we can express the optimal cut-off value for the energy efficiency maximization case with rate constraints using  (\ref{Eqn:EEWFPricingRateConstraintMacro}) as
\ifCLASSOPTIONonecolumn
\begin{align} \Omega_{EE,P,RC}^{\ast (n) } =   \left(\frac{\log(2)}{\Delta_f} \cdot \left(\lambda_{s} \cdot \Delta_M   + \mu_s \right) + \sum\limits_{j \neq k, j \in \mathcal{K}^{(n)}}  (1 + \tau_j)\pi_{k,j}^{(n)}\right)/\left( 1 + \tau_k \right), \label{Eqn:OmegaEERateConstraints}\end{align}
\else
\begin{align}\resizebox{\linewidth}{!}{$\Omega_{EE,P,RC}^{\ast (n) } =   \left(\frac{\log(2)}{\Delta_f} \cdot \left(\lambda_{s} \cdot \Delta_M   + \mu_s \right) + \sum\limits_{j \neq k, j \in \mathcal{K}^{(n)}}  (1 + \tau_j)\pi_{k,j}^{(n)}\right)/\left( 1 + \tau_k \right),$} \label{Eqn:OmegaEERateConstraints}\end{align}
\fi
where the initials RC stand for the rate constraints. Notice that when all the rate constraints are satisfied, the optimal cut-off value in (\ref{Eqn:OmegaEERateConstraints}) reduces to (\ref{Eqn:OmegaEE}).

We need to emphasize that this type of formulation enables us to satisfy two contradicting objectives of maximizing the average energy efficiency (or similarly, the aggregate throughput) and introducing the fairness among users. For example, users who are subject to high interference conditions or low channel gains are typically allocated low power levels due to the water-filling principle. For this reason, their throughput values  are typically low. The formulation in (\ref{Eqn:WFEEFracRateConstraints}) solves this problem by increasing their power levels through the Lagrangian variables associated with the minimum rate requirements. Thus, it ensures that the system fairness is increased.

\ifCLASSOPTIONonecolumn
\begin{algorithm}[h!]
\caption{Iterative Water-Filling Algorithm with Pricing for Network Energy Efficiency Maximization with Minimum Rate Constraints in Two-Tier Heterogeneous Networks}\label{Algorithm:EEWFAlgPricingTwoTierQoS}
      \begin{algorithmic}[1]
	\State Set the initial transmit power levels, interference prices, and dual prices, and initialize $t = 0$. At each sector, solve
 	\While{$|F_s(\lambda_{s})| > \epsilon$  and $l < l_{\max}$} 
    \State Determine $\lambda_s$ using the following
    \begin{align} \lambda_s(\textbf{p}_s,\textbf{p}_p) = \bigg(\sum_{n \in \mathcal{N}} \Delta_f \log_2(1 + p_s^{(n)} \chi_k^{(n)}) + \sum_{p \in \mathcal{S}_{P,s}} \sum_{n \in \mathcal{N}}\Delta_f \log_2 (1+ p_p^{(n)} \chi_k^{(n)}) \bigg)/\psi_s(\textbf{p}_s,\left\{ \textbf{p}_p \right\})  \end{align}
	\State Obtain $\mu_{s}$ using Algorithm~\ref{Algorithm:Bisection}
    \State For all $n \in \mathcal{N}$, solve for $p_{\text{Next}}^{(n)}$ using (\ref{Eqn:EEWFPricingRateConstraintMacro})
	\For{ all $p \in \mathcal{S}_{P,s}$}
	\State Obtain $\mu_{p}$ using Algorithm~\ref{Algorithm:Bisection}
    \State Solve for $p_{\text{Next,p}}^{(n)}$ for all $n \in \mathcal{N}$ using (\ref{Eqn:EEWFPricingRateConstraintPico})
	\EndFor                
	\State Update the dual prices, $\tau_k$ for all $k \in \mathcal{K}_s$, using (\ref{Eqn:EEWFDualPriceUpdate})
    \State Calculate the following
    \begin{align} F_s(\lambda_{s}) = \sum_{n \in \mathcal{N}} \Delta_f \log_2\left(1 + p_s^{(n)} \chi_k^{(n)}\right) + \sum_{p \in \mathcal{S}_{P,s}} \sum_{n \in \mathcal{N}}\Delta_f \log_2\left(1+ p_p^{(n)} \chi_k^{(n)}\right) - \lambda_s \psi_s(\textbf{p}_s,\left\{ \textbf{p}_p \right\}) \end{align}
	\State Update $l = l + 1$
	\EndWhile{}
	\State Update the power levels using
    \begin{align}\begin{aligned} p_{s}^{(n)}(t + 1) =& (1-\delta) \cdot p_{s}^{(n)}(t) + \delta \cdot p_{\text{Next}}^{(n)}  \text{ and }\\
    p_{p}^{(n)}(t + 1) =& (1-\delta) \cdot p_{p}^{(n)}(t) + \delta \cdot p_{\text{Next,p}}^{(n)} \text{ for all } p \in \mathcal{S}_{P,s}
    \end{aligned}
    \end{align}
    \State Distribute the interference prices, $\{ \pi_{k,j}^{(n)}\}$, among base stations
    \State Go to Step~\ref{Step:EEMax} and repeat for $t=t+1$
\end{algorithmic}
\end{algorithm}
\else
\begin{algorithm}[h!]
\caption{Iterative Water-Filling Algorithm with Pricing for Network Energy Efficiency Maximization with Minimum Rate Constraints in Two-Tier Heterogeneous Networks}\label{Algorithm:EEWFAlgPricingTwoTierQoS}
      \begin{algorithmic}[1]
      	\State Set the initial transmit power levels, interference prices, and dual prices, and initialize $t = 0$. At each sector, solve
 	\While{$|F_s(\lambda_{s})| > \epsilon$  and $l < l_{\max}$} 
    \State Determine $\lambda_s$ using the following
    \begin{align} \lambda_s =& \bigg(\sum_{n \in \mathcal{N}} \Delta_f \log_2(1 + p_s^{(n)} \chi_k^{(n)}) \\ & \hspace{-0.25em} + \sum_{p \in \mathcal{S}_{P,s}} \sum_{n \in \mathcal{N}}\Delta_f \log_2 (1+ p_p^{(n)} \chi_k^{(n)}) \bigg)/\psi_s(\textbf{p}_s,\left\{ \textbf{p}_p \right\})  \nonumber \end{align}
	\State Obtain $\mu_{s}$ using Algorithm~\ref{Algorithm:Bisection}
    \State For all $n \in \mathcal{N}$, solve for $p_{\text{Next}}^{(n)}$ using (\ref{Eqn:EEWFPricingRateConstraintMacro})
	\For{ all $p \in \mathcal{S}_{P,s}$}
	\State Obtain $\mu_{p}$ using Algorithm~\ref{Algorithm:Bisection}
    \State Solve for $p_{\text{Next,p}}^{(n)}$ for all $n \in \mathcal{N}$ using (\ref{Eqn:EEWFPricingRateConstraintPico})
	\EndFor                
\State Update the dual prices, $\tau_k$ for all $k \in \mathcal{K}_s$, using (\ref{Eqn:EEWFDualPriceUpdate})
   \State Calculate the following
   \begin{align} & F_s(\lambda_{s}) = \sum_{n \in \mathcal{N}} \Delta_f \log_2\left(1 + p_s^{(n)} \chi_k^{(n)}\right) \\ &  + \sum_{p \in \mathcal{S}_{P,s}} \sum_{n \in \mathcal{N}}\Delta_f \log_2\left(1+ p_p^{(n)} \chi_k^{(n)}\right) - \lambda_s \psi_s(\textbf{p}_s,\left\{ \textbf{p}_p \right\}) \nonumber \end{align}
	\State Update $l = l + 1$
	\EndWhile{}
	\State Update the power levels using
    \begin{align}\begin{aligned} p_{s}^{(n)}(t + 1) =& (1-\delta) \cdot p_{s}^{(n)}(t) + \delta \cdot p_{\text{Next}}^{(n)}  \text{ and } \nonumber \\
    p_{p}^{(n)}(t + 1) =& (1-\delta) \cdot p_{p}^{(n)}(t) + \delta \cdot p_{\text{Next,p}}^{(n)} \text{ for all } p \in \mathcal{S}_{P,s} \nonumber
    \end{aligned}
    \end{align}
    \State Distribute the interference prices, $\{ \pi_{k,j}^{(n)}\}$, among base stations
    \State Go to Step~\ref{Step:EEMax} and repeat for $t=t+1$
\end{algorithmic}
\end{algorithm}
\fi


The iterative water-filling algorithm for network energy efficiency maximization problem with minimum rate constraints in two-tier networks is given under the heading Algorithm~\ref{Algorithm:EEWFAlgPricingTwoTierQoS}. The dual prices associated with the minimum rate constraints are updated using 
\begin{align}\tau_{k}^{(l+1)} = \left[\tau_{k}^{(l)} - \alpha^{(l)} \left(\sum_{n \in \mathcal{N}_k} r_k^{(n)} - R_{\min,k} \right)\right]^+,\label{Eqn:EEWFDualPriceUpdate}\end{align}
where the operator $[x]^+$ denotes $\max(0,x)$. The step size at $l$th iteration is denoted by $\alpha^{(l)}$. We employ an adaptive step size selection algorithm such that \cite{BertsekasConstrained}
\ifCLASSOPTIONonecolumn
\begin{align}
\alpha^{(l)} = \begin{cases} \beta \alpha^{(l-1)} & \text{ if }  \left( R_{\min,k} - \sum\limits_{n \in \mathcal{N}_k} r_k^{(n,l)} \right) > \kappa \left( R_{\min,k} - \sum\limits_{n \in \mathcal{N}_k} r_k^{(n,l-1)}\right)  \\ 
\alpha^{(l-1)} & \text{ otherwise},
\end{cases} 
\end{align}
\else
\begin{align} 
\alpha^{(l)} = \begin{cases} \beta \alpha^{(l-1)} & \text{ if }  ( R_{\min,k} - \sum\limits_{n \in \mathcal{N}_k} r_k^{(n,l)}) >  \\
& \hspace{5em} \kappa ( R_{\min,k} - \sum\limits_{n \in \mathcal{N}_k} r_k^{(n,l-1)}) \\
\alpha^{(l-1)} & \text{ otherwise},
\end{cases} 
\end{align}
\fi
where $r_k^{(n,l)}$ and $r_k^{(n,l-1)}$ are the throughput of user~$k$ on subcarrier~$n$ at iterations $l$ and $l-1$, respectively. The scalar $\beta$ increases the step size if the difference between the minimum rate requirement and the throughput of a user is not decreased by a factor of $\kappa$ in the next time instant. In the simulations, we take the step size as $\alpha^{(0)} = 2.5\times 10^{-4}$, the increment factor $\beta$ as $2$, and the comparison threshold $\kappa$ as $0.9$. This step size rule is studied more in detail in \cite[p.~123]{BertsekasConstrained} to update the dual prices in constrained optimization problems. The proposed algorithm starts transmitting at an initial transmit power. Dual prices and interference prices are taken as zero initially. The algorithm calculates $\lambda_s$, and using this value, power levels for the macrocell and picocell base stations are determined. We update the dual prices and repeat this process until the convergence criterion is satisfied. To avoid rapid fluctuations, we use the Mann iterations as in Algorithm~\ref{Algorithm:EEWFAlgPricing}. Finally, interference prices are measured at the user and these measurements are fed back to the base stations, where they are distributed among base stations using the fast and reliable backhaul (for example, through the X2-interface in LTE, see \cite{holma}), and the process is repeated in the next time slot.


\section{Simulation Results}\label{Section:EEWFSimResults}
In this section we present the simulation results for the single-tier and two-tier energy-efficiency maximization problems. In the simulation model, we follow the simulation models and parameters suggested in \cite{36814} as a baseline simulation for LTE heterogeneous networks. We consider a network consisting of 19 hexagonal macrocell deployments and each macrocell has 3-sector antennas. In each sector, 30~users are randomly generated within the macrocell sector area and each user is equipped with a single omni-directional antenna. This corresponds to the uniform user distribution scenario in \cite{36814}. For the two-tier simulation model, we deploy four picocells per sector. We consider a non-uniform user distribution where two users are initially dropped within a 40~meter radius per picocell base station and the remaining users are randomly generated. This model is also proposed in \cite{36814}. We adopt the same simulation parameters and models as in \cite{CCC2014}. For the scheduler, we employ the Equal Bandwidth Scheduler which is detailed in \cite{36814,DavasliScheduling,CCC2014}. For spectrum allocation, we consider the fractional frequency reuse scheme in \cite{CCC2014}, which is shown to achieve very high energy efficiency performance in two-tier heterogeneous networks compared to other benchmark spectrum allocations. We will investigate two problems: energy efficiency and throughput maximization, and for each problem we consider the non-pricing and pricing scenarios. For the macrocell base stations, $P_{\text{Total,s}} = 46$~dBm and $P_{\text{Total,p}} = 30$~dBm \cite{36814}. Also, for simplicity, we take $P_{\max,s}^{(n)}$ and $P_{\max,p}^{(n)}$ as zero. The base station power consumption model parameters are taken as $P_{0,m} = 130$~W, $P_{0,p} = 56$~W, $\Delta_M = 4.7$, and $\Delta_P = 2.6$ as in \cite{FettweisChristian12}. Note that when a picocell base station has no associated users, we consider that it is in dormant mode and it consumes $P_{\text{Sleep,P}} = 6.3$~W.


\begin{figure}[t!]
\centering
\ifCLASSOPTIONonecolumn
		\begin{tabular}{cc}
		\subfigure[]{\includegraphics[width=0.5\columnwidth]{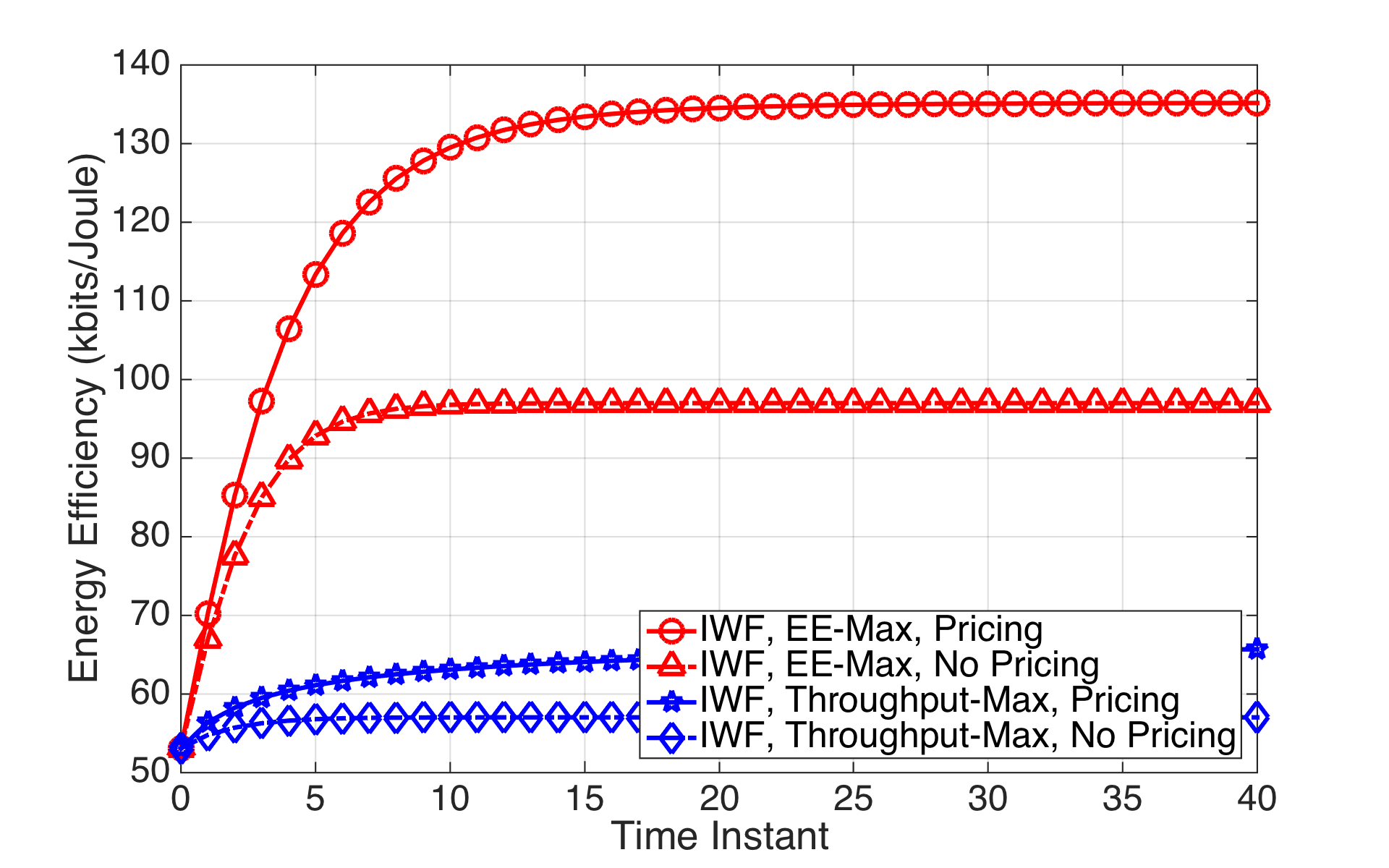}}&
		\subfigure[]{\includegraphics[width=0.5\columnwidth]{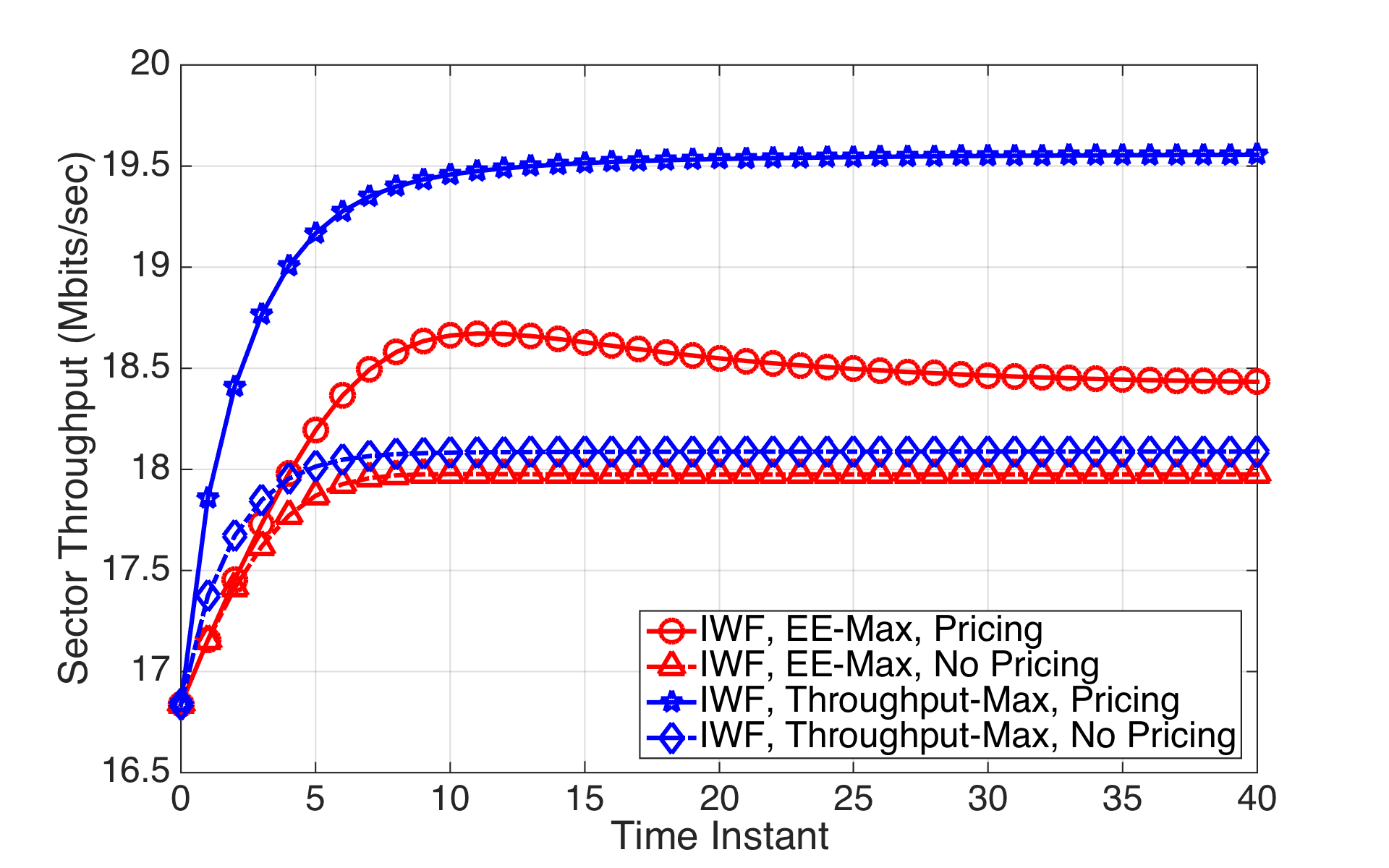}}\\						
		\subfigure[]{\includegraphics[width=0.5\columnwidth]{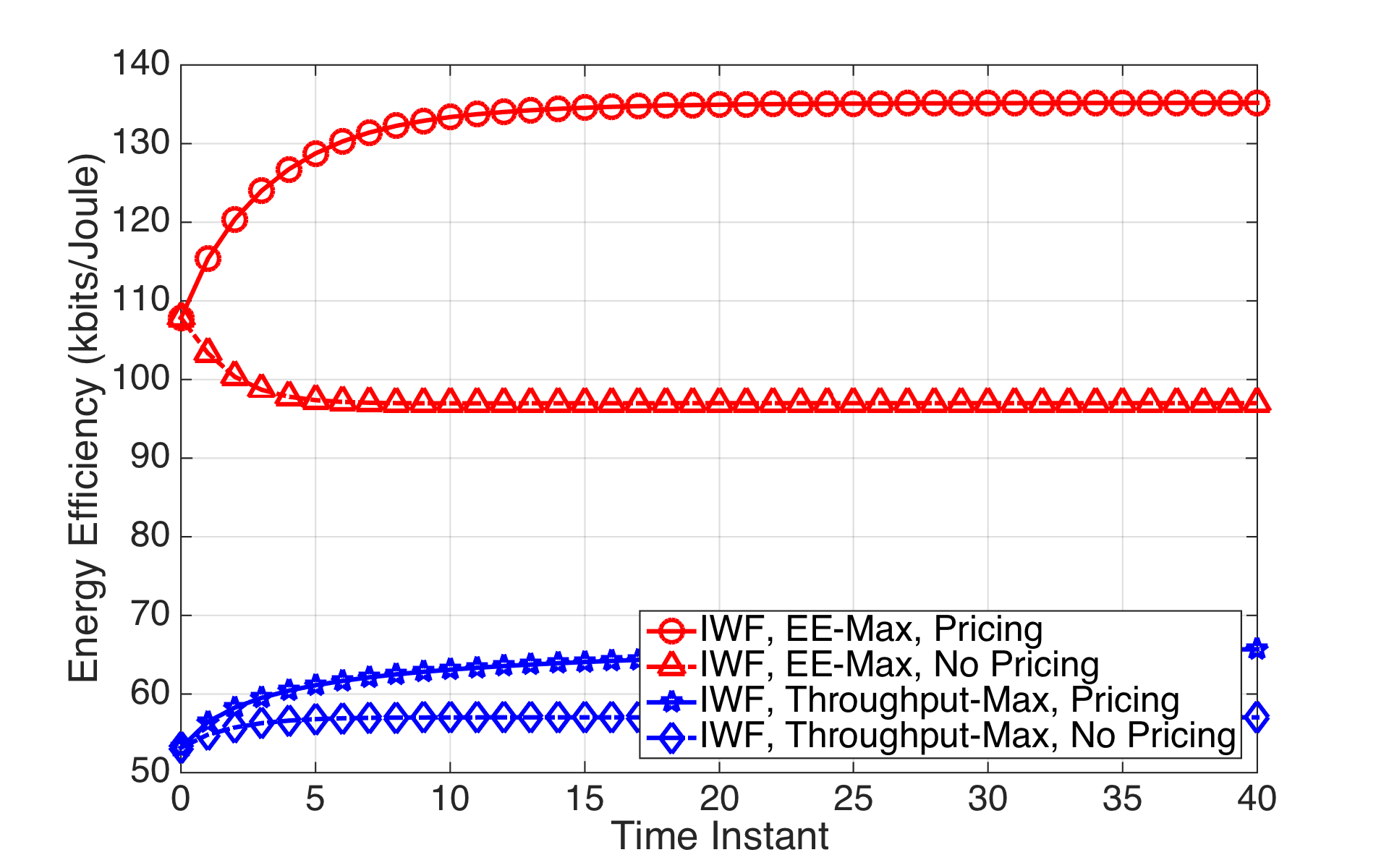}}&
		\subfigure[]{\includegraphics[width=0.5\columnwidth]{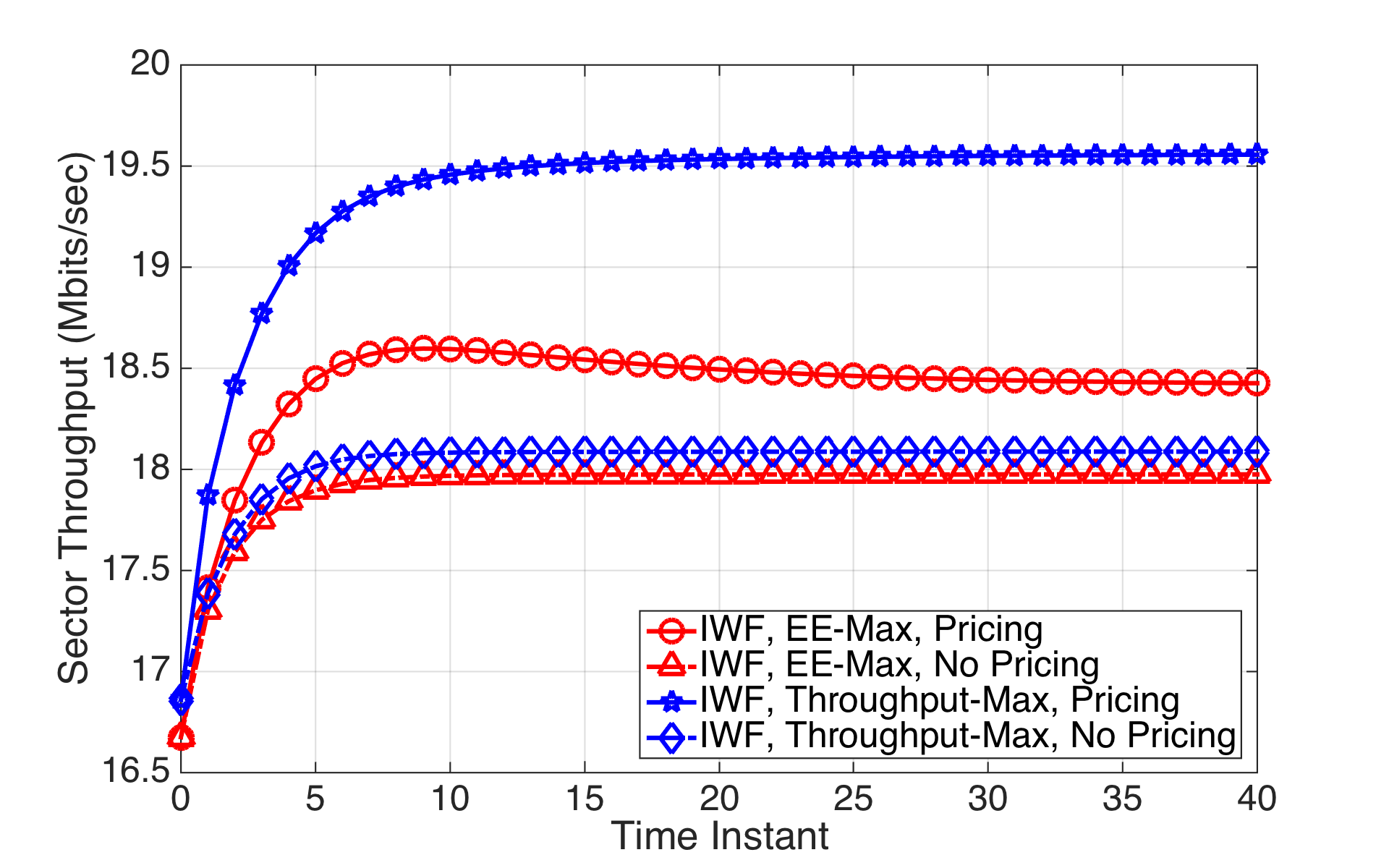}}\\
	\end{tabular}
\else
		\begin{tabular}{c}
		\subfigure[]{\includegraphics[width=0.7\columnwidth]{EE_WF_All_Warmup1in10_SL_30Aug-31-2015_17-38-46M1M2M3M4.png}}\\
		\subfigure[]{\includegraphics[width=0.7\columnwidth]{CAP_WF_All_Warmup1in10_SL_0_Users30Aug-31-2015_17-38-46M1M2M3M4.png}}\\						
		\subfigure[]{\includegraphics[width=0.7\columnwidth]{EE_WF_All_NoWarmup1in10_SL_30Aug-31-2015_17-38-46M1M2M3M4.png}}\\
		\subfigure[]{\includegraphics[width=0.7\columnwidth]{CAP_WF_All_NoWarmup_SL_0_Users30Aug-31-2015_17-38-46M1M2M3M4.png}}
		\end{tabular}
\fi
\caption{Average sector energy efficiency and throughput of a single-tier network using the proposed iterative water-filling algorithms with different initial power levels. IWF stands for iterative water-filling. The solutions without pricing correspond to the algorithm in \cite{FettweisChristian12}.}
\label{Figure:EECapWFWarmupSingleLayer}
\end{figure}

\subsection{Results in Single-Tier Networks}
In Figs.~\ref{Figure:EECapWFWarmupSingleLayer}(a)-(d), we present the average sector energy efficiency and throughput results for a single-tier network. These four figures investigate different initial power levels for warm-up. In Figs.~\ref{Figure:EECapWFWarmupSingleLayer}(a)-(b), we start the simulations with initially transmitting at maximum power levels, whereas the power levels are determined without any interference price information Figs.~\ref{Figure:EECapWFWarmupSingleLayer}(c)-(d). It can be observed that both initial power levels converge to the same point after $40$~time instants. Also, we observe that power control improves the energy efficiency and throughput by factors of $2.53$ and $1.10$ for the energy efficiency maximization problem, respectively, and $22\%$ in energy efficiency and $16\%$ in throughput for the throughput maximization problem. 

When we compare the resource allocation with and without interference pricing, we observe that interference pricing brings $40\%$ and $13\%$  additional improvements in terms of energy efficiency for the energy efficiency and throughput maximization problems, respectively. Note that the scenario without interference pricing corresponds the algorithm proposed in \cite{FettweisChristian12}. Hence, it can be concluded that the proposed algorithm outperforms the one in \cite{FettweisChristian12}. 

\begin{figure}[t!]
\centering
	\ifCLASSOPTIONonecolumn
		\includegraphics[width=0.565\columnwidth]{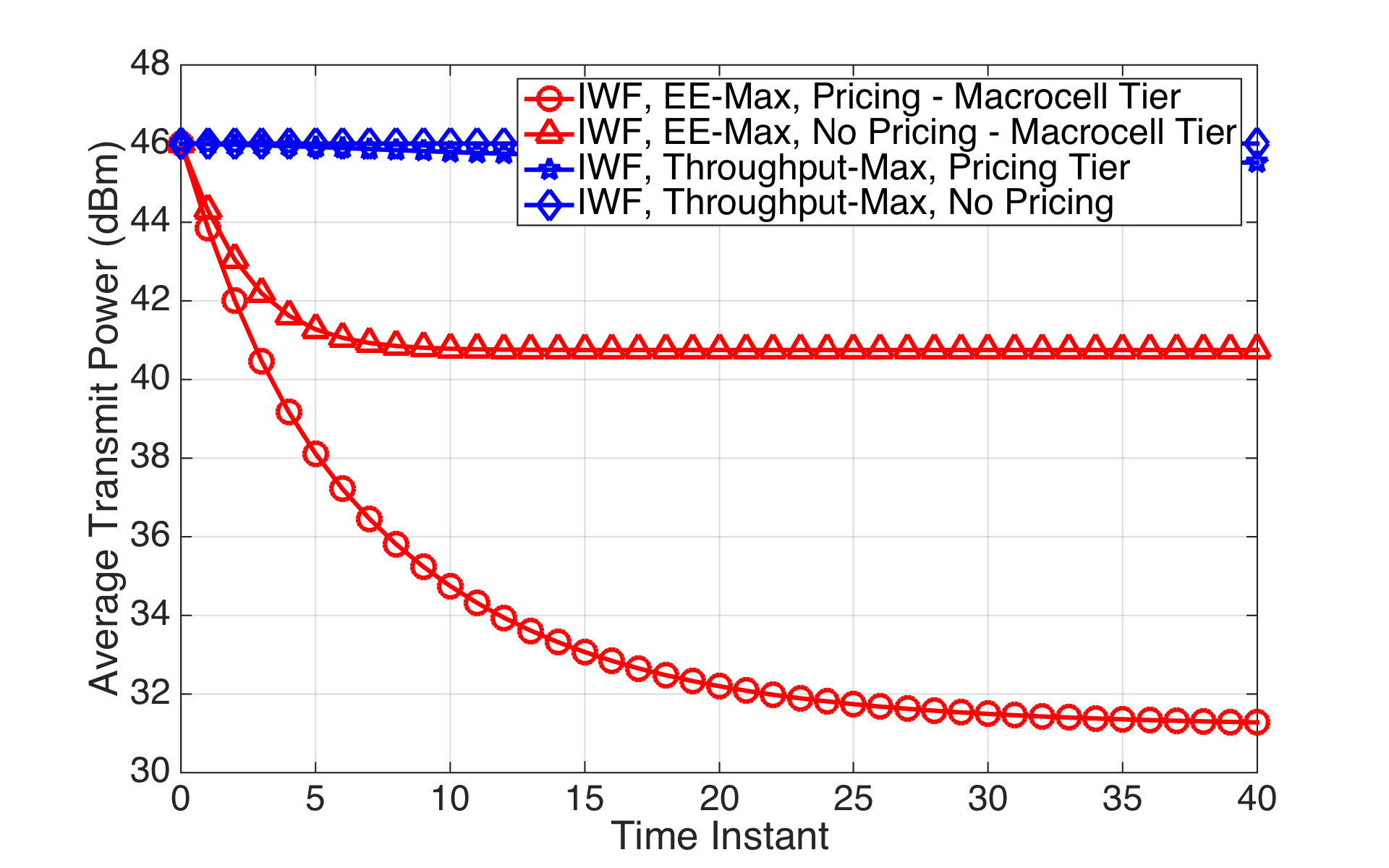}
	\else
		\includegraphics[width=0.7\columnwidth]{Power_All_Warmup1in10_SL_0_Users30Aug-31-2015_17-38-46M1M2M3M4}
\fi
\caption{Average power consumption of the proposed iterative water-filling algorithms in a single-tier network.}
\label{Figure:PowerWFWarmupSingleLayer}
\end{figure}

\begin{figure}[t!]
\centering
	\ifCLASSOPTIONonecolumn
	\begin{tabular}{c}		
	\subfigure[]{\includegraphics[width=0.565\columnwidth]{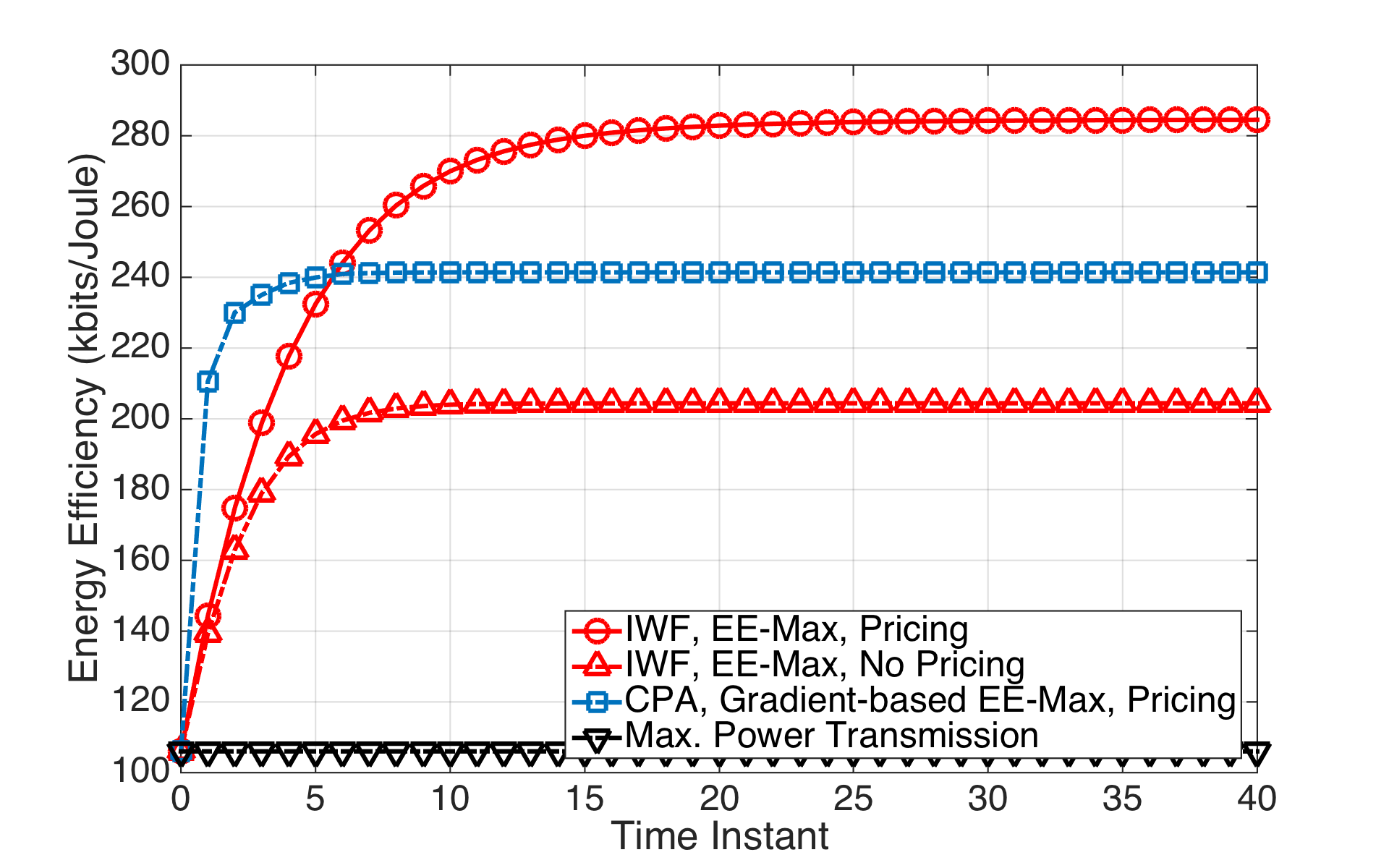}}\\
	\subfigure[]{\includegraphics[width=0.565\columnwidth]{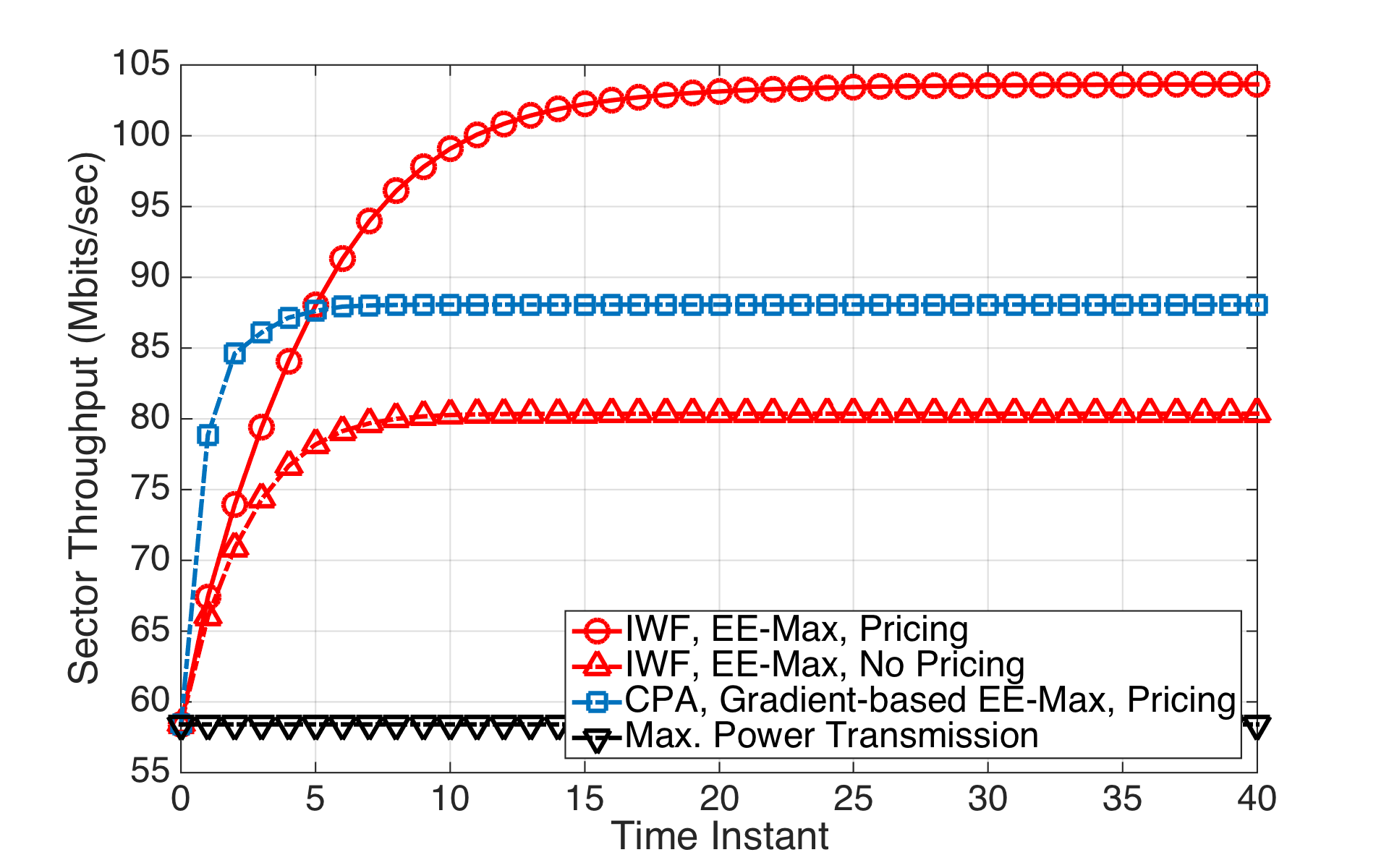}}	
	\end{tabular}
	\else
	\begin{tabular}{c}		
	\subfigure[]{\includegraphics[width=0.7\columnwidth]{EE_WF_All_Warmup1in10_HetNet_4_Users30Aug-31-2015_21-51-25M1M1M1M1}}\\
	\subfigure[]{\includegraphics[width=0.7\columnwidth]{CAP_WF_All_Warmup1in10_HetNet_4_Users30Aug-31-2015_21-51-25M1M1M1M1}}	
	\end{tabular}
\fi
\caption{Average sector energy efficiency and throughput of a two-tier network using the proposed iterative water-filling algorithms. CPA corresponds to the constant power allocation algorithm proposed in \cite{CCC2014}.}
\label{Figure:EECapWFWarmupTwoLayer}
\end{figure}
\begin{figure}[t!]
\centering
	\ifCLASSOPTIONonecolumn
		\includegraphics[width=0.5\columnwidth]{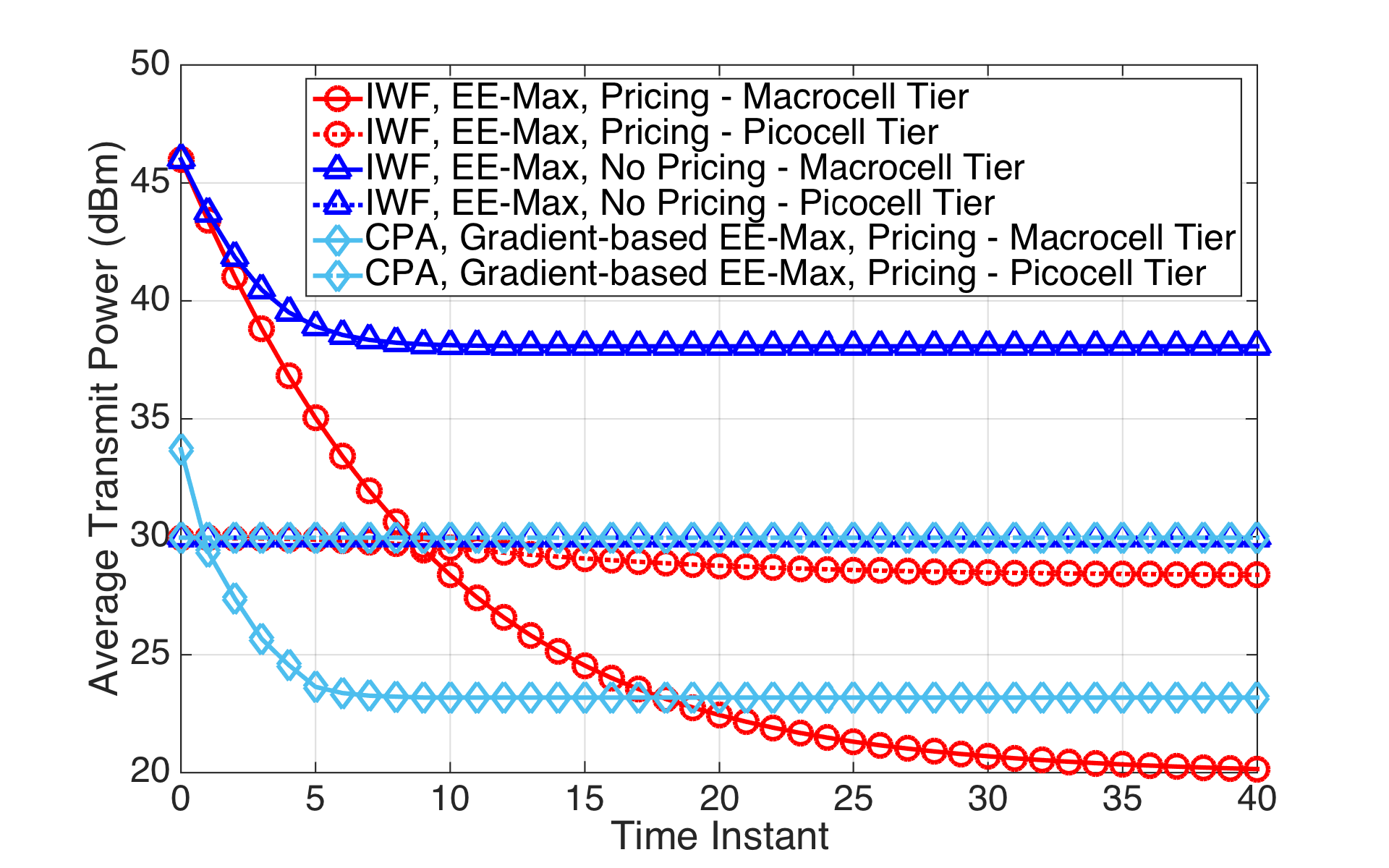}
	\else
		\includegraphics[width=0.7\columnwidth]{Power_WF_All_Warmup1in10_HetNet_4_Users30Aug-31-2015_21-51-25M1M1M1M1}
	\fi
	\caption{Average transmit power consumption of the proposed iterative water-filling algorithms with and without pricing in a two-tier network.}
\label{Figure:PowerWFWarmupTwoLayer}
\end{figure}

Figure~\ref{Figure:PowerWFWarmupSingleLayer} illustrates another advantage of the proposed algorithm: It brings significant power savings. When we apply power control without interference pricing, the average transmit power reduces from $39.8$~W ($46$~dBm) to $11.89$~W ($40.75$~dBm), which corresponds to a reduction of $3.35$ times. It is worth noting that when base stations communicate among each other to exchange interference prices, it can bring additional power savings. For example, in the energy efficiency maximization problem, average transmit power of macrocell base station reduces from $11.89$~W ($40.75$~dBm) to $1.66$~W ($32.20$~dBm) when pricing is introduced. Thus, we observe that interference pricing brings a power reduction of $7$ times compared to the case without pricing and $24$ times compared to the case without power control, which are very significant. 



\subsection{Results in Two-Tier Networks}
Figures~\ref{Figure:EECapWFWarmupTwoLayer}(a)-(b) depict the average energy efficiency and aggregate sector throughput for the iterative water-filling algorithm with and without pricing in two-tier heterogeneous networks. Note again that the case without pricing corresponds to the algorithm in \cite{FettweisChristian12}. Also, for comparison, we evaluate the performance of the maximum power case and constant power allocation with pricing which was proposed in \cite{CCC2014}. We observe that power control improves the energy efficiency and throughput by factors of $2.68$ and $1.77$, respectively. Interference pricing brings $39\%$ improvement in energy efficiency and $29\%$ in throughput over the case without pricing.

Figure~\ref{Figure:PowerWFWarmupTwoLayer} presents the transmit power consumption of each tier using the above algorithms. First, we observe that significant power savings can be achieved in the macrocell tier, whereas picocell base stations typically operate close to the maximum power levels. For example, in the case with pricing, iterative water-filling  algorithm reduces the power consumption from the initial maximum power level of $46$~dBm down to $20.2$~dBm, whereas the average transmit power of a picocell base station is slightly reduced from $30$~dBm to $28.4$~dBm. Also, pricing mechanism brings an additional $3.6$ times average transmit power saving per sector compared to the case without pricing, reducing it from $40.2$~dBm to $34.6$~dBm, which is very significant. These results illustrate why picocells should be deployed as an underlying tier such that users can be offloaded from the macrocell tier to the small cell tiers where the link distances are smaller and higher rates can be achieved.



\begin{figure}[t!]
\centering
	\ifCLASSOPTIONonecolumn
	\begin{tabular}{c}	
	\subfigure[]{\includegraphics[width=0.6\columnwidth]{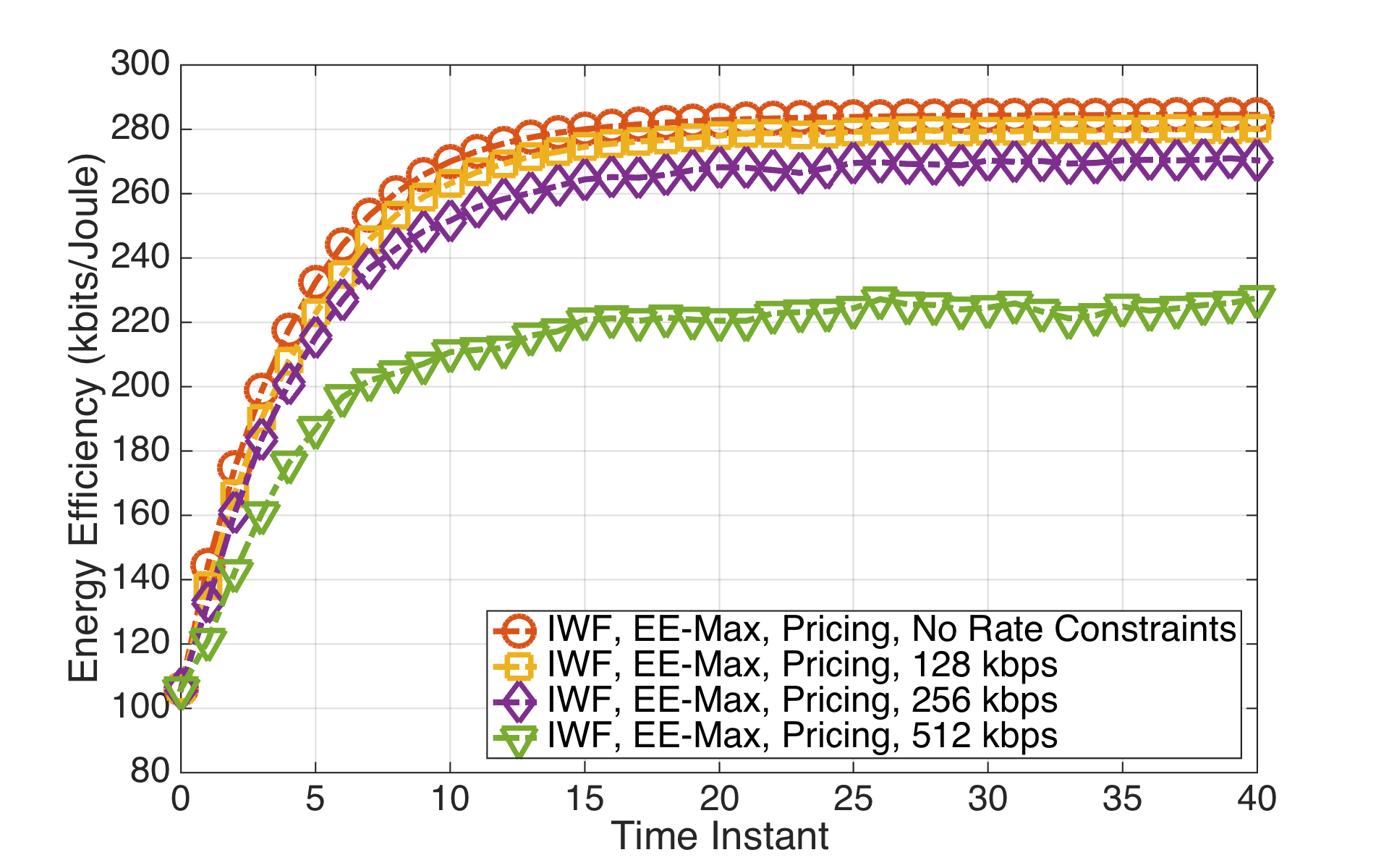}}\\
	\subfigure[]{\includegraphics[width=0.6\columnwidth]{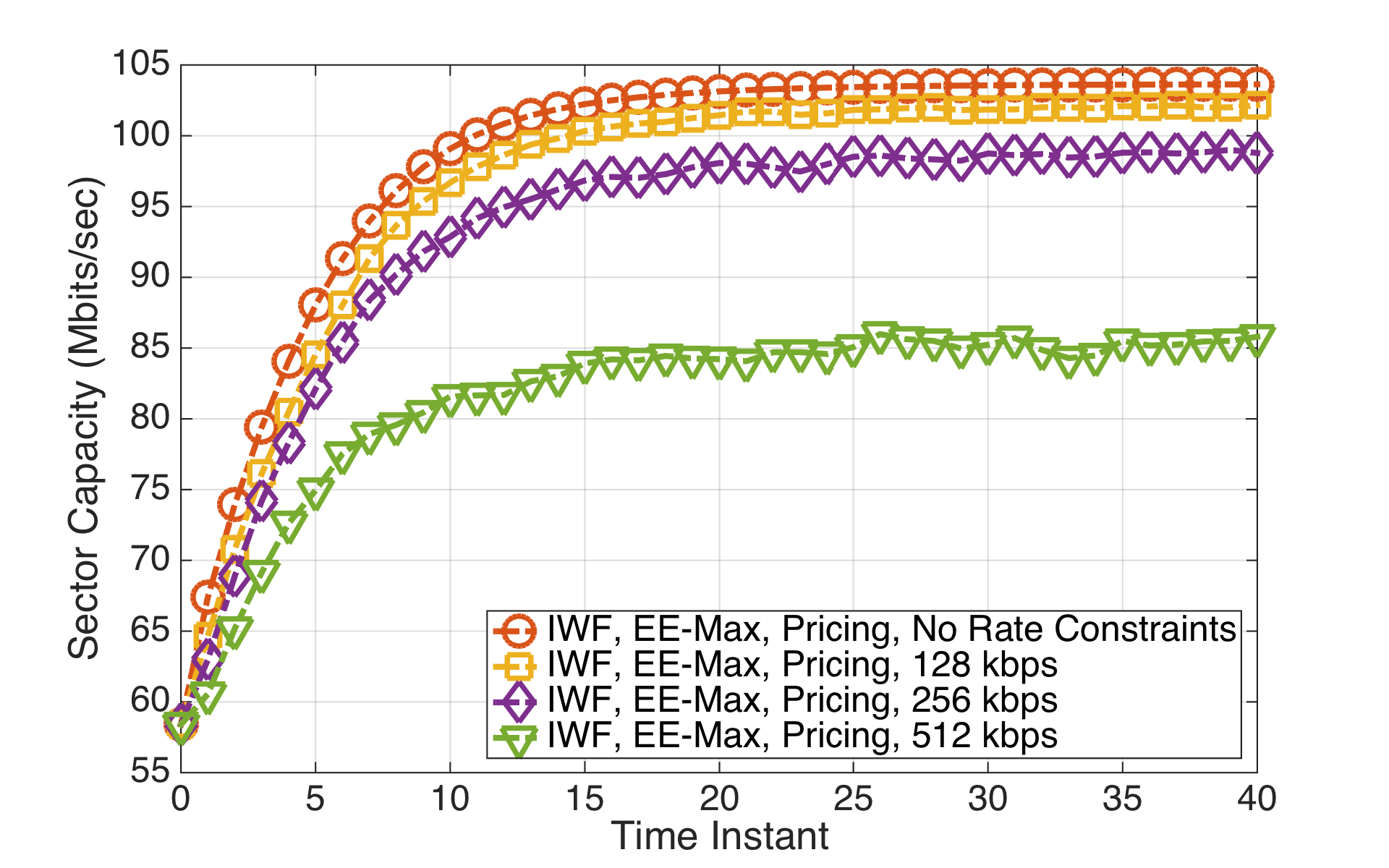}}
	\end{tabular}
	\else
	\begin{tabular}{c}
	\subfigure[]{\includegraphics[width=0.7\columnwidth]{EE_WF_All_QoS_Warmup1in10_HetNet_4_Users30Sep-08-2015_17-37-02M1}}\\
	\subfigure[]{\includegraphics[width=0.7\columnwidth]{CAP_WF_All_QoS_Warmup1in10_HetNet_4_Users30Sep-08-2015_17-37-02M1}}
	\end{tabular}
	\fi
\caption{Average sector energy efficiency and sector throughput for various minimum rate requirements.}
\label{Figure:EEWFQoSEECap}
\end{figure}

\begin{figure}[th!]
\centering
\ifCLASSOPTIONonecolumn
	\begin{tabular}{c}
	\subfigure[]{\includegraphics[width=0.6\columnwidth]{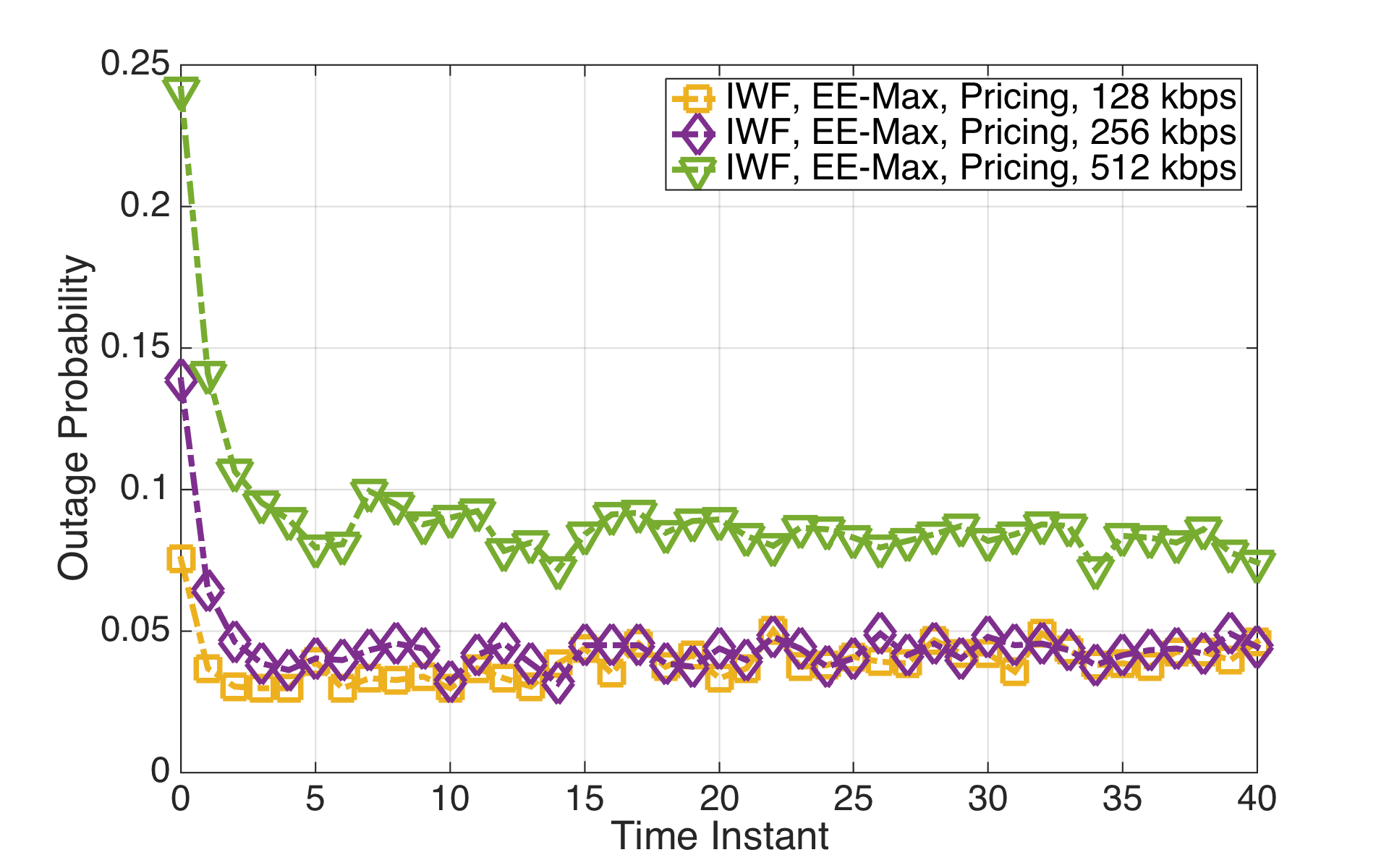}}\\
	\subfigure[]{\includegraphics[width=0.6\columnwidth]{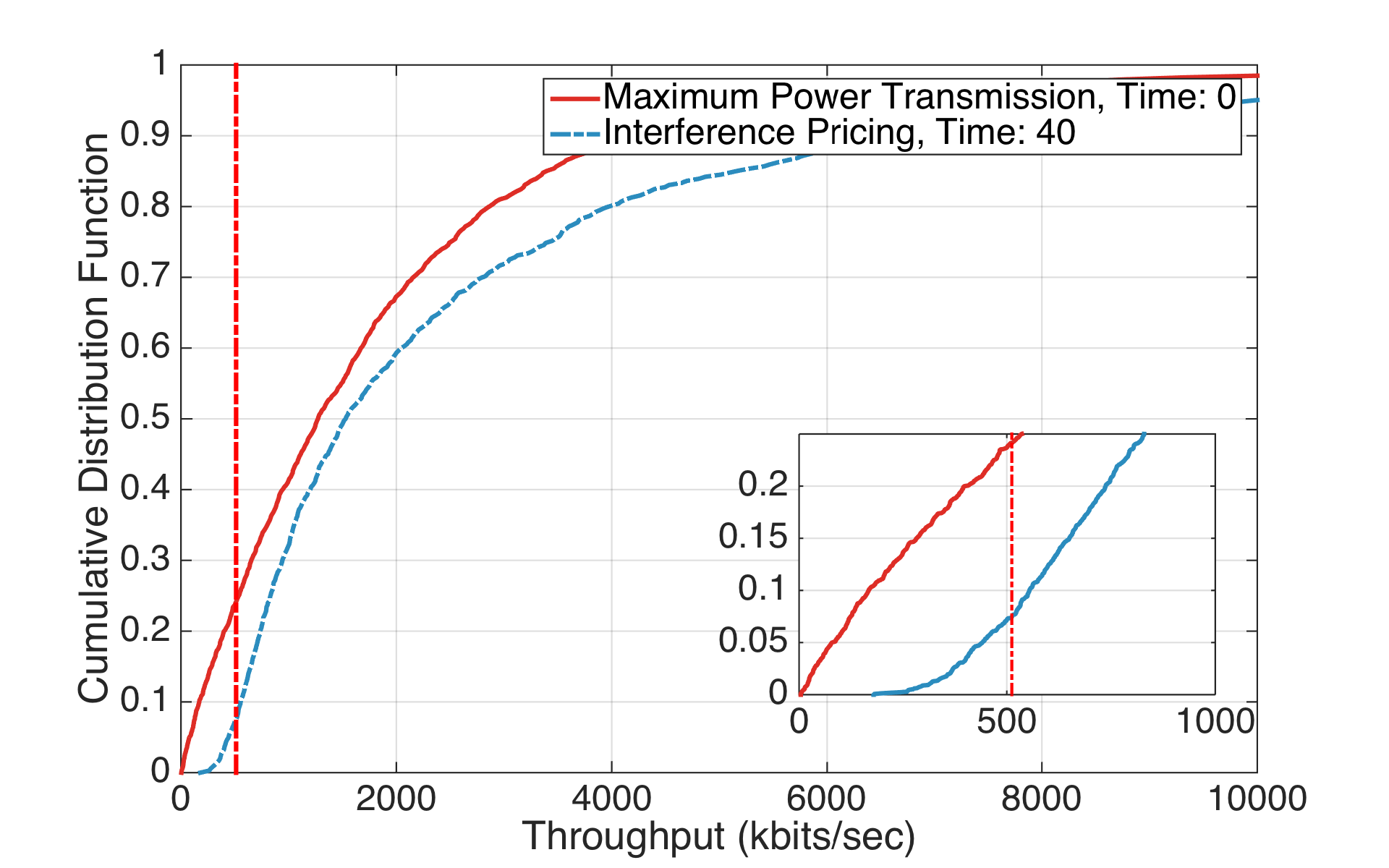}}	
	\end{tabular}
\else
	\begin{tabular}{c}
	\subfigure[]{\includegraphics[width=0.7\columnwidth]{Outage_WF_All_QoS_Warmup1in10_HetNet_4_Users30Sep-08-2015_17-37-02M1}}\\
	\subfigure[]{\includegraphics[width=0.7\columnwidth]{RateCDF_WF_All_QoS_Warmup1in10_HetNet_4_Users30Sep-08-2015_17-37-02M1}}
	\end{tabular}
\fi
\caption{(a) The outage probability of various minimum rate requirements and (b) the cumulative distribution function of user rates for the minimum rate requirement of $512$~kbits/sec.}
\label{Figure:EEWFQoSOutageRateCDF}
\end{figure}

\subsection{Minimum Rate Constraints in Two-Tier Networks}
Finally, we extend the iterative water-filling algorithm for the energy efficiency maximization problem and we incorporate the minimum rate constraints. For simplicity, the same target rate is considered for all users. We need to note that, in real applications, users may have different rate requirements. For example, \cite{holma} considers a mixture of different traffic requirements consisting of best-effort users and users with strict rate requirements. Fig.~\ref{Figure:EEWFQoSEECap} illustrates the average energy efficiency and sector throughput performance of the proposed algorithm for minimum rate requirements ranging from $128$~kbits/sec up to $512$~kbits/sec. First, we observe that as the rate requirement increases, the average sector energy efficiency decreases. As we have derived in (\ref{Eqn:OmegaEERateConstraints}), the rate requirements are enforced through adjusting the water-filling levels. However, this comes at the expense of reductions in energy efficiency and throughput. For example, the average energy efficiency is $284.5$~kbits/Joule without any rate constraints and it reduces to $187.6$~kbits/Joule for rate constraints of $512$~kbits/sec. When the rate requirements are not satisfied, the users are considered to be in outage. Fig.~\ref{Figure:EEWFQoSOutageRateCDF}(a) presents the outage probability of users for different minimum rate requirements. As expected, a higher rate requirement yields a higher outage probability. As the dual prices are updated and interference prices are distributed, the number of users in outage decreases significantly. For example, when power control is not employed for the $512$~kbits/sec case, the outage probability is $25\%$. Using the proposed algorithm, the outage probability gradually decreases to $7\%$ at the end of $40$~iterations. Fig.~\ref{Figure:EEWFQoSOutageRateCDF}(b) illustrates the cumulative distribution of user rates. It  depicts how the user rate distribution is improved using the proposed algorithm. We observe that the proposed algorithm outperforms the case without power control, shifting every point of the cumulative distribution to the right.

\section{Conclusions}\label{Section:EEWFQoSOutage}
Resource allocation in multi-cell networks is an important aspect for cellular wireless systems. In this paper, we investigated the energy efficiency maximization problem from a power control perspective. We considered a realistic load-adaptive base station power consumption model capturing the characteristics of a macrocell and a picocell base station. We  obtained closed-form expressions for the water-filling solutions using methods from fractional programming. We  proposed several iterative water-filling algorithms for LTE networks with single-tier and two-tier deployments. We incorporated interference pricing mechanism in which base stations communicate among themselves to exchange limited information. Then, the preceding framework was extended to incorporate the minimum rate constraints per user. The corresponding closed-form expressions for the case with minimum rate constraints were derived as well. The average energy efficiency, throughput, and transmit power consumption performance of the proposed algorithms were evaluated and compared to other baseline works. The numerical results demonstrated that the proposed algorithms can achieve significant gains and outperform the baseline methods.

\bibliographystyle{IEEEtran}
\bibliography{IEEEabrv,AdaptiveFFR2}
\end{document}